\ifx\onecol\undefined
\documentclass[journal,twocolumn]{IEEEtran}
\else 
\documentclass[12pt,draftclsnofoot,journal,onecolumn]{IEEEtran}
\fi

\IEEEoverridecommandlockouts

\usepackage{geometry}
\usepackage[noend]{algpseudocode}
\usepackage{balance}
\usepackage{algorithmicx,algorithm}
\usepackage{eqparbox}
\usepackage{amsmath}
\usepackage{subfigure}
\usepackage{verbatim}
\usepackage{braket}
\usepackage{cases}
\usepackage{amsthm,amsmath,amssymb,bm,bbm}

\newtheorem{lemma}{Lemma}
\newtheorem{theorem}{Theorem}
\newtheorem{remark}{Remark}
\newtheorem{corollary}{Corollary}
\usepackage{mathrsfs}

\usepackage{hyperref}
\hypersetup{hidelinks, 
colorlinks=true,
allcolors=black,
pdfstartview=Fit,
breaklinks=true}

\usepackage{graphicx}
\usepackage{epstopdf}
\usepackage{cite}
\makeatletter
\newcommand{\biggg}{\bBigg@{3}}
\def\bigggl{\mathopen\biggg}

\newcommand{\Biggg}{\bBigg@{3.5}}

\newcommand{\bigggg}{\bBigg@{4}}

\newcommand{\Bigggg}{\bBigg@{4.5}}

\makeatother

\begin{document}
%
%
%
%


\title{Can Continuous Aperture MIMO Achieve Much Better Performance than Discrete MIMO?
}
\author{
		\IEEEauthorblockN{
		Zhongzhichao~Wan,
		Jieao~Zhu,
		and~Linglong~Dai,~\IEEEmembership{Fellow,~IEEE}
	}
	\thanks{All authors are with the Department of Electronic Engineering, Tsinghua University as well as Beijing National Research Center for Information Science and Technology (BNRist), Beijing 100084, China (E-mails: \{wzzc20, zja21\}@mails.tsinghua.edu.cn; daill@tsinghua.edu.cn).}
	\thanks{This work was supported in part by the National Key Research and Development Program of China (Grant No. 2020YFB1807201), in part by the National Natural Science Foundation of China (Grant No. 62031019).}

}

\maketitle



%

\begin{abstract}
	The concept of continuous-aperture multiple-input multiple-output (CAP-MIMO) technology has been proposed recently, which aims at achieving high spectrum efficiency by deploying extremely dense antennas or even continuous antennas in a given aperture. The fundamental question of CAP-MIMO is whether it can achieve much better performance than the traditional discrete MIMO system. In this paper, to model the CAP-MIMO, we use self-adjoint operators to depict the structural characteristics of the continuous random electromagnetic fields from physical laws. Then, we propose a non-asymptotic performance comparison scheme between continuous and discrete MIMO systems based on the analysis of mutual information. We show the consistency of the proposed scheme by proving that the mutual information between discretized transceivers converges to that between continuous transceivers. Numerical analysis verifies the theoretical results, and suggests that the mutual information obtained from the discrete MIMO with widely adopted half-wavelength spaced antennas almost achieves the mutual information obtained from CAP-MIMO. 
\end{abstract}


\section{Introduction}

The spectrum efficiency of wireless communication systems has been greatly improved from 3G to 5G because of the use of multiple-input multiple-output (MIMO) technology \cite{Jeffrey'14,Federico'14,Andrews'16}. The MIMO systems utilize multiple antennas to exploit the spatial multiplexing gain \cite{Geoffrey'14}, where the antennas are modeled as discrete points in the continuous space. Along with the tendency of increasing the number of antennas to achieve higher spectrum efficiency, people are considering deploying extremely dense antennas in a given aperture \cite{decarli2021communication, zhang2022pdma}. When the number of antennas in a given aperture tends to infinity, the traditional MIMO systems with transceivers composed of discrete point antennas are equivalent to the MIMO systems with continuously controllable transceivers.
Therefore, the MIMO with extremely dense antennas is called continuous-aperture MIMO (CAP-MIMO), and is also called holographic MIMO \cite{Demir'21,Marzetta'20,sanguinetti2022wavenumber} or large intelligent surface \cite{decarli2021communication,Jide'20} in the recent literature\footnote{The MIMO with extremely dense antennas can be accurately described by the name CAP-MIMO, while holographic MIMO and large intelligent surface do not focus on the continuity of the transceiver apertures. Therefore, in the rest part of the paper, we will prefer using the name CAP-MIMO rather than using other names like holographic MIMO.}. It has attracted increasing interest in the research of MIMO technology. Recent works about CAP-MIMO include pattern optimization \cite{zhang2022pdma}, antenna design \cite{Yurduseven'18}, channel estimation \cite{Demir'21}, and so on. 
For CAP-MIMO, the fundamental question is whether the CAP-MIMO system can achieve much better performance than the traditional discrete MIMO system.   

\subsection{Related works}
The structure of CAP-MIMO has been defined in the previous part but there are many structures for realizing the discrete MIMO. Therefore, we need to choose which structure of the discrete MIMO to compare with CAP-MIMO. A representative structure of discrete MIMO uses half-wavelength spaced antennas to compose the transceivers \cite{Yiwei'20,zhang2022beam,wu2019intelligent}, because half-wavelength sampling of the electromagnetic field can reconstruct the original field according to the sampling theorem. 

There have been several works discussing the performance
comparison between CAP-MIMO and discrete MIMO with 
half-wavelength spaced antennas. The performance comparison is from the degrees of freedom (DoF) perspective. Specifically, when discarding the evanescent wave components, the Fourier transform of the received field, which is in the wavenumber domain, is concentrated in a circle or a segment. This concentration phenomenon means that the field is bandlimited in the wavenumber domain, 
thus it can be perfectly recovered from the half-wavelength sampling points in the spatial domain \cite{migliore2018near} according to the Nyquist sampling theorem \cite{landau1967sampling}. 
The above conclusion is based on the assumption that we can observe the received field in the {\it infinitely} large spatial domain. However, in practice, the destination where we can observe the field is in a {\it finitely} large aperture. 

For a rigorous analysis framework of the DoF in a {\it finitely} large aperture, the prolate spheroidal wave function (PSWF) \cite{slepian1961prolate} is introduced to perform orthogonal expansion on the electromagnetic field. Specifically, to reconstruct the wavenumber-bandlimited electromagnetic field observed in a length-$l$ spatial region, the PSWFs were used as the basis based on the Slepian's concentration problem \cite{slepian1976bandwidth}. Such an electromagnetic field can be perfectly reconstructed from infinite number of PSWFs, and approximately reconstructed from a finite number of PSWFs. If the reconstruction error can be controlled within a given threshold by using $N_0$ PSWFs, the number of DoFs of the field can be approximated by $N_0$ \cite{wavethoeyofinformation}. This analyzing scheme is strict for arbitrary $l$, 
but can only provide the {\bf asymptotic} result of the DoF, i.e., the quantitative result of $N_0$ can be obtained only when the length $l$ or the frequency tends to infinity. Under this assumption, $N_0$ equals the number of antennas under half-wavelength sampling.
However, the practical systems are with {\it finitely} large aperture and finite frequency. The {\bf asymptotic} result can not provide quantitative number of DoFs for practical systems. Therefore, a {\bf non-asymptotic} performance comparison scheme between CAP-MIMO and discrete MIMO is required for the accurate performance comparison with {\it finitely} large apertures. 

\subsection{Our contributions}
To solve this problem, in this paper, we provide a {\bf non-asymptotic} performance comparison scheme between CAP-MIMO and discrete MIMO, and we further prove the rationality of the scheme\footnote{Simulation codes will be provided to reproduce the results in this paper: \url{http://oa.ee.tsinghua.edu.cn/dailinglong/publications/publications.html}.}. 
Specifically, the contributions of this paper can be summarized as follows:

\begin{itemize}
	\item{We build models of CAP-MIMO and discrete MIMO based on electromagnetic theory. For CAP-MIMO with continuous transceivers, we model the structural characteristics of the continuous random electromagnetic fields from physical laws by using self-adjoint operators. Based on this model, we can utilize the spectrum theory of operators to derive the information that can be obtained from the received field. The existing models of MIMO with discrete transceivers are spatially discretized from the continuous model. Moreover, signal-to-noise ratio (SNR) control schemes are introduced to ensure the fairness of the comparison between CAP-MIMO and discrete MIMO.}
	\item {\begin{sloppypar}Then, before comparing the performance between CAP-MIMO with continuous transceivers and traditional MIMO with discrete {\it transceivers}, we first utilize the simplified model with continuous transmitter and discrete {\it receiver}. Under this simplified model, the transmitter is continuous, which is the same as that in the CAP-MIMO system. By theoretically analyzing the mutual information that can be obtained from the discrete receiver in this simplified model, we can obtain some insights about how the discretization of the receiver affects the mutual information. Moreover, the theoretical proof of the convergence of the mutual information in the simplified model can inspire the analysis of a more practical scenario, i.e., the discrete {\it transceivers}. \end{sloppypar} 	
	}
	\item{Finally, we extend the convergence proof from the model with discrete {\it receiver} to the model with discrete {\it transceiver}. We prove that the mutual information between the discrete transceivers converges to the mutual information between continuous transceivers when the number of antennas of the discretized transceivers tends to infinity. Therefore, the continuous model in our paper is compatible with the existing discrete models, and the fairness of the performance comparison is guaranteed. Numerical results are provided to verify the theoretical analysis. Moreover, it shows the near-optimality of the half-wavelength sampling of the transceivers in traditional discrete MIMO.}
\end{itemize}

\subsection{Organization and notation}
\emph{Organization}: The rest of our paper is organized as follows. Section.~\ref{sec_models} introduces the basic model of EIT and proposes models with continuous or discrete transceivers. The mutual information between the transceivers is also derived. Section.~\ref{sec_discrete_receiver} proves the convergence of the mutual information between continuous transmitter and discrete receiver when the number of discrete antennas increases. 
Then, the convergence of the mutual information between discrete transceivers is illustrated in Section \ref{sec_discrete_transceiver}. Finally, we conclude the paper in Section \ref{sec_conclusion}.

\emph{Notation}: bold characters denote matrices and vectors; ${\rm j}$ is the imaginary unit;
${\mathbb E}\left[x\right]$ denotes the mean of random variable $x$; $x^{*}$ denotes the conjugation of a number or a function $x$; ${\bf X}^{\rm H}$ denotes the conjugate transpose of a vector or a matrix ${\bf X}$; 
$\mu_0$ is the permeability of a vacuum, $Z_0$ is the free-space intrinsic impedance and $c$ is the speed of light in a vacuum; 
$\nabla$ is the nabla operator, and $\nabla \times$ is the curl operator; 
$\ket{\phi}$ is the quantum mechanical notation of a function $\phi$, where the inner product is denoted by $\bra{\psi}\phi\rangle$;
$\det(\cdot)$ denotes the matrix determinant or the Fredholm determinant; ${\rm tr}(\cdot)$ denotes the trace of a matrix or an operator. ${\bf I}_m$ denotes the $m \times m$ identity matrix, ${\bf 1}$ denotes the identity operator, $\delta(x)$ denotes the delta function, and ${\mathbbm 1}_{i=j}$ denotes the indicator function; $|x|$ denotes the modulus of a complex variable, and $\left\| f(x) \right\|_{L^{\infty}(a,b)}$ is the uniform norm of the function $f(x)$ over the interval $[a,b]$. $C^\infty(K)$ denotes the set of smooth functions supported on a compact set $K$. 

\section{Models of continuous and discrete systems}
\label{sec_models}
In this section, we introduce the models of continuous and discrete systems for performance comparison between CAP-MIMO and discrete MIMO. We control the SNR at the receiver side to ensure the fairness of the comparison. The information obtained from these models is derived from operators and matrices.  
\subsection{Basic model of electromagnetic information theory}
To model the transceivers and the channel, we follow the approach of electromagnetic information theory (EIT). The EIT is an interdisciplinary subject that integrates the classical electromagnetic theory and information theory to build an analysis framework for the ultimate performance bound of wireless communication systems \cite{zhu2022electromagnetic}. The analysis framework of EIT is based on spatially continuous electromagnetic fields, which provides us the tool to model and analyze the continuous transceivers. Then, for the consistency, the model of discrete transceivers are viewed as the discretization of the continuous model from EIT.

The model of EIT is built on the vector wave equation~\cite{dardari2020communicating} without boundary conditions, which is expressed by 
\begin{equation}
\nabla  \times \nabla  \times {\bf{E}}\left( {\bf{r}} \right) - {\kappa_0 ^2}{\bf{E}}\left( {\bf{r}} \right) = {\rm j}\omega {\mu _0}{\bf{J}}\left( {\bf{r}} \right) = {\rm j}\kappa_0 {Z_0}{\bf{J}}\left( {\bf{r}} \right),
\label{VWE}
\end{equation}
where $\kappa_0  = \omega \sqrt {\mu_0 \varepsilon_0 }$ is the wavenumber, and $Z_{0}=\mu_0 c = 120 \pi\,{\rm  [\Omega]}$ is the free-space intrinsic impedance. 

We assume that the transceivers are confined in two regions $V_{\rm s}$ and $V_{\rm r}$, separately. 
The current density at the source is ${\bf J}({\bf s})$, where ${\bf s}\in\mathbb{R}^3$ is the coordinate of the source. The induced electric field at the destination is ${\bf E}({\bf r})$, where ${\bf r} \in \mathbb{R}^3$ is the coordinate of the field observer. 
To solve the linear partial differential equation~\eqref{VWE}, a general theoretical approach is to introduce the dyadic Green's function ${\bf{G}} ({\bf{r}},{\bf{s}})\in \mathbb{C}^{3\times 3}$.
According to the linearity of \eqref{VWE}, the electric field ${\bf{E}}({\bf{r}})$ can be expressed by  
\begin{equation}
	{\bf{E}}({\bf{r}}) = \int_{{V_s}} {\bf{G}} ({\bf{r}},{\bf{s}}){\bf{J}}({\bf{s}}){\rm d}{\bf{s}},\quad {\bf{r}} \in {V_r}.
	\label{VWE_Green}
\end{equation}
By exploiting the symmetric properties of the free space, the Green's function in unbounded, homogeneous mediums at a fixed frequency point is~\cite{poon2005degrees}
\ifx\onecol\undefined 
	\begin{equation}
		\begin{aligned}
			{\bf{G}}({\bf{r}},{\bf{s}}) &= \frac{{\rm j}\kappa_0 {Z_0}}{{4\pi }} \left( {{\bf{I}} + \frac{{{\nabla _{\bf{r}}}\nabla _{\bf{r}}^{\rm{H}}}}{{{\kappa_0 ^2}}}} \right) \frac{{{e^{{\rm{j}}\kappa_0 \left\| {{\bf{r}} - {\bf{s}}} \right\|}}}}{{\left\| {{\bf{r}} - {\bf{s}}} \right\|}}  \\
			&= \frac{{\rm j}\kappa_0 {Z_0}}{{4\pi }}\frac{{{e^{{\rm{j}}\kappa_0 \left\| {{\bf{r}} - {\bf{s}}} \right\|}}}}{{\left\| {{\bf{r}} - {\bf{s}}} \right\|}}\Bigg[\left( {{\bf{I}} - {\bf{\hat p}}{{{\bf{\hat p}}}^{\rm{H}}}} \right) \\&~~~+ \frac{{\rm j}}{2\pi \left\| {{\bf{r}} - {\bf{s}}} \right\| /\lambda}\left( {\bf I}-3{\bf{\hat p}}{{{\bf{\hat p}}}^{\rm{H}}} \right) \\&~~~-\frac{1}{(2\pi\left\| {{\bf{r}} - {\bf{s}}} \right\|/\lambda )^2 }\left( {\bf I}-3{\bf{\hat p}}{{{\bf{\hat p}}}^{\rm{H}}}  \right) \Bigg] [{\rm \Omega}/{\rm m}^2],
			\label{Green}
		\end{aligned}
	\end{equation}
\else 
	\begin{equation}
		\begin{aligned}
			{\bf{G}}({\bf{r}},{\bf{s}}) &= \frac{{\rm j}\kappa_0 {Z_0}}{{4\pi }} \left( {{\bf{I}} + \frac{{{\nabla _{\bf{r}}}\nabla _{\bf{r}}^{\rm{H}}}}{{{\kappa_0 ^2}}}} \right) \frac{{{e^{{\rm{j}}\kappa_0 \left\| {{\bf{r}} - {\bf{s}}} \right\|}}}}{{\left\| {{\bf{r}} - {\bf{s}}} \right\|}}  
		{\bf{G}}({\bf{r}},{\bf{s}})  
			= \frac{{\rm j}\kappa_0 {Z_0}}{{4\pi }}\frac{{{e^{{\rm{j}}\kappa_0 \left\| {{\bf{r}} - {\bf{s}}} \right\|}}}}{{\left\| {{\bf{r}} - {\bf{s}}} \right\|}}\Bigg[\left( {{\bf{I}} - {\bf{\hat p}}{{{\bf{\hat p}}}^{\rm{H}}}} \right) \\&~~+ \frac{{\rm j}}{2\pi \left\| {{\bf{r}} - {\bf{s}}} \right\| /\lambda}\left( {\bf I}-3{\bf{\hat p}}{{{\bf{\hat p}}}^{\rm{H}}} \right) -\frac{1}{(2\pi\left\| {{\bf{r}} - {\bf{s}}} \right\|/\lambda )^2 }\left( {\bf I}-3{\bf{\hat p}}{{{\bf{\hat p}}}^{\rm{H}}}  \right) \Bigg] [{\rm \Omega}/{\rm m}^2],
			\label{Green}
		\end{aligned}
	\end{equation}
\fi

where ${\bf{\hat p}} = \frac{{\bf{p}}}{{\left\| {\bf{p}} \right\|}}$ and ${\bf{p}} = {\bf{r}} - {\bf{s}}$. 

Since there are some non-ideal factors at the receiver that corrupts the received field, we call them the noise field ${\bf N}({\bf r})$. The received electric field can be expressed by ${\bf Y}({\bf r})={\bf E}({\bf r})+{\bf N}({\bf r})$.
 The above equations represent the deterministic model in the electromagnetic theory. To satisfy the demand of wireless communication, we need to convey information through the electromagnetic field. 
Specifically, the wireless communication system encodes the information in the current ${\bf J}({\bf s})$, and decodes the information from the noisy electric field ${\bf Y}({\bf r})$. Due to the randomness of the transmitted bit source, the electromagnetic fields are randomly excited by the transmitter equipment before being radiated into the propagation media. Therefore, the electromagnetic fields should be modeled as random fields~\cite{yaglom1957some}, and we adopt the Gaussian random fields to depict the statistical characteristics of the fields. We denote the autocorrelation function of the current and the electric field as matrix-valued functions ${\bf R}_{\bf J}({\bf s},{\bf s}^{'}) = \mathbb{E}[{\bf J}({\bf s}){\bf J}^{\rm H}({\bf s}^{'})]$ and ${\bf R}_{\bf E}({\bf r},{\bf r}^{'}) = \mathbb{E}[{\bf E}({\bf r}){\bf E}^{\rm H}({\bf r}^{'})]$. The relationship between ${\bf R}_{\bf J}$ and ${\bf R}_{\bf E}$ is determined by the Green's function, which is
\begin{equation}
	\begin{aligned}
	{\bf R}_{\bf{E}}({\bf{r}},{\bf{r}}') 
	=\int_{V_s}\int_{V_s}{\bf G}({\bf r},{\bf s}){\bf R}_{\bf{J}}({\bf{s}},{\bf{s}}'){\bf G}^{\rm H}({\bf r},{\bf s}){\rm d}{\bf s}{\rm d}{\bf s}'.
	\end{aligned}
	\end{equation}
Similar definitions of the autocorrelation functions for the noise field and the noisy electric field are represented as ${\bf R}_{\bf N}({\bf r},{\bf r}') = \mathbb{E}[{\bf N}({\bf r}){\bf N}^{\rm H}({\bf r}')]$ and ${\bf R}_{\bf Y}({\bf r},{\bf r}') = \mathbb{E}[{\bf Y}({\bf r}){\bf Y}^{\rm H}({\bf r}')]$.

\subsection{Continuous transceivers}
\label{subsec_continuous transceivers}
In this part, we will build the model of CAP-MIMO with continuous transceivers based on the EIT model in the above subsection, and then derive the mutual information between the continuous transceivers. 
For simplicity, in the rest part of the paper, we assume that the transceivers are linear along the ${\hat{z}}$-direction. Moreover, since the current $J$ can only exist on the linear source and we only observe the electric field on the linear receiver, we express all the physical quantities in a Cartesian coordinate system that satisfies ${\bf s} = (0,0,s)$ and ${\bf r} = (d,0,r)$, where $d$ is the distance between the parallel source and destination line. This model corresponds to single-polarized linear antennas. Through this simplification scheme, we use $J(s)$ and $E(r)$ instead of ${\bf J}(\bf s)$ and ${\bf E}(\bf r)$. The relationship between them can be expressed by 
	$E(r) = \int_0^l G(r,s)J(s){\rm d}s$,
where $G(r,s)$ is the bottom-right element of the matrix ${\bf G}({\bf r},{\bf s})$, i.e., $G={\bf G}_{3,3}$. We can derive $G(r,s)$ as
\ifx\onecol\undefined
\begin{equation}
	\begin{aligned}
	\label{equ:Green_g}
	{G}{(r,s)} =& \frac{{{\rm j} {Z_0} {e^{{\rm j}2\pi \sqrt {{x^2} + {d^2}} /\lambda }}}}{{2\lambda \sqrt {{x^2} + {d^2}} }}\Big[ \frac{\rm j}{{2\pi \sqrt {{x^2} + {d^2}} /\lambda }}\frac{{{d^2} - 2{x^2}}}{{{x^2} + {d^2}}} 
	\\&+\frac{{{d^2}}}{{{x^2} + {d^2}}}
	- \frac{1}{{{{(2\pi /\lambda )}^2}({x^2} + {d^2})}}\frac{{{d^2} - 2{x^2}}}{{{x^2} + {d^2}}}\Big],
	\end{aligned}
	\end{equation} 
	\else 
	\begin{equation}
		\begin{aligned}
		\label{equ:Green_g}
		{G}{(r,s)} =& \frac{{{\rm j} {Z_0} {e^{{\rm j}2\pi \sqrt {{x^2} + {d^2}} /\lambda }}}}{{2\lambda \sqrt {{x^2} + {d^2}} }}\Big[ \frac{\rm j}{{2\pi \sqrt {{x^2} + {d^2}} /\lambda }}\frac{{{d^2} - 2{x^2}}}{{{x^2} + {d^2}}} 
		+\frac{{{d^2}}}{{{x^2} + {d^2}}}
		- \frac{1}{{{{(2\pi /\lambda )}^2}({x^2} + {d^2})}}\frac{{{d^2} - 2{x^2}}}{{{x^2} + {d^2}}}\Big],
		\end{aligned}
		\end{equation} 
	\fi
	where $x = r-s$ and $\lambda = 2\pi/\kappa_0$ is the wavelength.

Here we consider the scenario with no channel state information, which means that the signals on the source are under equal power allocation. 
The second moments (autocorrelation) of $J$ are denoted by $R_J(s,s') = P\delta(s-s'), s,s'\in [0,l]$. 

Since the noiseless received field is uniquely determined by the source and the deterministic channel, the autocorrelation function of the electric field is expressed by the source autocorrelation $R_J(s,s')$ and the Green's function $G(r,s)$, written as
\ifx\onecol\undefined
\begin{equation}
	\begin{aligned}
	R_E(r,r') &= \int_{0}^{l} \int_{0}^{l} G(r,s)R_J(s,s') G^{*}(r',s'){\rm d}s{\rm d}s' \\&= P\int_{0}^{l} G(r,s)G^{*}(r',s){\rm d}s.
	\end{aligned}
	\label{equ:R_E_noCSI}
\end{equation}
\else
\begin{equation}
	\begin{aligned}
	R_E(r,r') &= \int_{0}^{l} \int_{0}^{l} G(r,s)R_J(s,s') G^{*}(r',s'){\rm d}s{\rm d}s' = P\int_{0}^{l} G(r,s)G^{*}(r',s){\rm d}s.
	\end{aligned}
	\label{equ:R_E_noCSI}
\end{equation}
\fi

 The received field on the destination is ${Y}(r) = {E}(r)+{N}(r)$, where ${N}(r)$ is the noise field at the receiver. In this paper, we consider thermal noise model ${\mathbb E}\left[ N(r)N^{*}(r') \right] = \frac{n_0}{2}\delta(r-r')$. According to~\cite{wan2022mutual}, we can perform Mercer expansion on the electric field $E(r)$ to obtain a set of mutually independent random variables $\xi_k$. The expansion can be written as $E(r) = \sum_k \xi_k \phi_k(r)$, where $\mathbb{E}[\xi_{k_i}\xi^{*}_{k_j}] = \lambda_{k_i} \mathbbm{1}_{i=j}$ and $\langle\phi_{k_i}(r),\phi_{k_j}(r) \rangle = \delta_{k_i k_j}$. This expansion scheme has split the continuous field into independent components. Since the white noise field can be expanded under arbitrary orthogonal bases, the continuous channel is also decomposed into independent subchannels, which makes the mutual information of the subchannels summable.

	Next we will show that for the operator $T_E := \phi(r)\rightarrow \int_0^l K_E(r,r')\phi(r'){\rm d}r'$, where $K_E(r,r') = R_E(r,r') = P\int_0^l G(r,s)G^*(r',s){\rm d}s$, all of its eigenvalues are real and nonnegative. Moreover, the sum of its eigenvalues $\sum_{i=1}^{\infty}\lambda_i$ equals $P\int_0^l\int_0^l G(r,s)G^*(r,s){\rm d}r{\rm d}s $. Notice that $T_E$ can be decomposed to $T^*T$, where $T:=\phi(r) \rightarrow \sqrt{P}\int_{0}^{l} G(r,s)\phi(r){\rm d}s$ and $T^{*}:=\phi(r) \rightarrow \sqrt{P}\int_{0}^{l} G^{*}(r,s)\phi(r){\rm d}s$.
	This decomposition means that $T_E = T_E^{*}$ is a self-adjoint operator. We assume that $\lambda$ is an eigenvalue of $T_E$ and $\phi(r)$ is the corresponding eigenfunction. Since
	\ifx\onecol\undefined
	\begin{equation}
		\begin{aligned}
			\lambda &= \lambda \langle \phi(r),\phi(r)\rangle = \langle T_E^{*}\phi(r),\phi(r) \rangle \\&= \langle \phi(r), T_E\phi(r)\rangle = {\lambda}^{*},
		\end{aligned}
	\end{equation}
	we know that $\lambda$ is real. From 
	\begin{equation}
		\begin{aligned}
			\lambda &= \lambda\langle\phi(r),\phi(r)\rangle = \langle T^*T\phi(r),\phi(r)\rangle \\&= \langle T\phi(r),T\phi(r) \rangle \geqslant 0,
		\end{aligned}
	\end{equation}	
	we know that $\lambda$ is nonnegative.
	\else
	\begin{equation}
		\begin{aligned}
			\lambda = \lambda \langle \phi(r),\phi(r)\rangle = \langle T_E^{*}\phi(r),\phi(r) \rangle = \langle \phi(r), T_E\phi(r)\rangle = {\lambda}^{*},
		\end{aligned}
	\end{equation}
	we know that $\lambda$ is real. From 
	\begin{equation}
		\begin{aligned}
			\lambda = \lambda\langle\phi(r),\phi(r)\rangle = \langle T^*T\phi(r),\phi(r)\rangle = \langle T\phi(r),T\phi(r) \rangle \geqslant 0,
		\end{aligned}
	\end{equation}	
	we know that $\lambda$ is nonnegative.
	\fi
	
	From \cite{lax2002functional} we know that an integral operator on $[a,b]$ is a trace class operator if its kernel $K(x,y)$ satisfies $K(x,y)$ and $\partial_yK(x,y)$ are continuous on $[a,b]^2$. Therefore $T_E$ is a trace class operator, which means that the sum of its eigenvalues is finite and can be expressed by~\cite{gohberg1978introduction}
	\begin{equation}
		\begin{aligned}
			{\rm tr}(T_E) &= \int_0^l K_E(r,r){\rm d}r = P\int_0^l\int_0^l
			 G(r,s)G^*(r,s){\rm d}r{\rm d}s.
		\end{aligned}
		\label{equ_trace_TE}
	\end{equation}			

\begin{corollary}
	The non-negative values $\frac{\lambda_k}{n_0/2}$ represent the SNR of the independent subchannels.
	The mutual information between the noisy received field and the current on the source can be expressed by
\begin{equation}
	\begin{aligned}
	I_0({J}; {Y})=\sum_{k=1}^{+\infty}{\rm log}\left(1+\frac{\lambda_k}{n_0/2}\right).
	\end{aligned}
	\label{mercer_mutual_information}
\end{equation}
\end{corollary}

By introducing the Fredholm determinant which is the determinant of operators \cite{simon2005trace}, we can express (\ref{mercer_mutual_information}) by
\begin{equation}
	\begin{aligned}
	I_0({J}; {Y})={\rm log}\,{\det}\left({\bf 1}+\frac{T_E}{n_0/2}\right),
	\end{aligned}
	\label{Fredholm_mutual_information}
\end{equation}
where $(T_E\phi)(r) := \int_{0}^{L}R_E(r,r')\phi(r'){\rm d}r'$ and $\lambda_k$ are the eigenvalues of $T_E$.

\begin{remark}
	Extension of the mutual information analysis: our analysis here is based on the simplified model with uni-polarized linear antennas as the transceiver. This simplification reduces the dimension of the problem, where random fields degenerate to one-dimensional random processes. For the more general scenarios, such analyzing schemes are still effective. If the random field is defined in a region $X$, we can expand ${\bf E}({\bf r})$ by ${\bf E}({\bf r}) = \sum_k \xi_k \boldsymbol{\Phi}_k({\bf r})$ and its autocorrelation function ${\bf R}_{\bf E}({\bf r},{\bf r}')$ by ${\bf R}_{\bf E}({\bf r},{\bf r}') = \sum_k \lambda_k \boldsymbol{\Phi}_k({\bf r}) \boldsymbol{\Phi}^{\rm H}_k({\bf r}')$ \cite{de2013extension}. The expansion satisfies that $\lambda_k$ and $\boldsymbol{\Phi}_k({\bf r})$ are eigenvalues and eigenfunctions of the integral equation $\int_{X} {\bf R}_{\bf E}({\bf r},{\bf r}')\boldsymbol{\Phi}({\bf r}'){\rm d}{\bf r}' = \lambda_k \boldsymbol{\Phi}({\bf r}) $. Similar expressions of the mutual information in~\eqref{mercer_mutual_information} and~\eqref{Fredholm_mutual_information} can be derived.
\end{remark}

\subsection{Continuous transmitter and discrete receiver}
\label{subsec_discrete_receiver}
Before building the model with discrete transceivers, in this subsection, we will first build a simplified model with continuous transmitter and discrete receiver. The simplified model analyzed here can bring some insights about the discretization of both transceivers and the SNR control schemes. For the continuous transmitter, we still use the length-$l$ linear transmitter along the $\hat{z}$-direction.  
For the discrete receiver, we build a model with $m$ point antennas on a segment parallel to the linear transmitter in the destination region. The $i_{\rm th}$ point antenna is placed on $r_i \in [0,l]$. The correlation matrices of the received signals and received noise are denoted by ${\bf K}_E^{'}$ and ${\bf K}_N^{'}$. For the received signals, we assume that it is the sampling of the continuous electric field on the point $r_i$, which means that ${\bf K}_E^{'} = K_E(r_i,r_j)$. However, for the received noise on the antenna, it can not directly be assumed as the point sampling of the noise field, because of the delta function. To solve this problem, we assume that ${\bf K}_N^{'} = \frac{n_1}{2}{\bf I}_m$ is an identity matrix, and control the signal-to-noise ratio (SNR) of this model the same as that of the continuous model to ensure the fairness of the comparison. The SNR at the receiver of the continuous model is $\sum_{i=1}^{\infty}\frac{\lambda_i}{n_0/2}$, where $\lambda_i$ is the $i_{\rm th} $ eigenvalue of the operator $T_E$. From {\bf Lemma 1} we know that $\sum_{i=1}^{\infty}\frac{\lambda_i}{n_0/2} = \frac{P}{n_0/2}\int_0^l\int_0^l G(r,s)G^*(r,s){\rm d}r{\rm d}s$ is finite. The SNR at the receiver of the discrete model is $\sum_{i=1}^{m}\frac{\lambda^{'}_{i}}{n_1/2}$, where $\lambda^{'}_{i}$ is the $i_{\rm th}$ eigenvalue of the matrix ${\bf K}_E^{'}$.  

The SNR control scheme is necessary because if we do not control the SNR, the mutual information that can be obtained from the discrete antennas in the receiver may infinitely increase. Let us take a counter-example where the power of received signal and received noise on each point antenna remain unchanged when the number of antennas in a given aperture increases. For dense antennas we can assume that $N$ received signals of the antennas in a small aperture are nearly the same, while the corresponding noises are independent according to the model. Then, the SNR for the $N$ antennas will keep near-linearity increasing with $N$, since when we perform combing of the $N$ received signals we have ${\rm SNR} = \frac{\mathbb{E}[(\sum_{i=1}^N E_i)(\sum_{i=1}^N E^{*}_i)]}{\mathbb{E}[(\sum_{i=1}^N N_i)(\sum_{i=1}^N N^{*}_i)]} \approx N \frac{\mathbb{E}[E_1 E^{*}_1]}{\mathbb{E}[N_1 N^{*}_1]}$. Therefore the mutual information that can be obtained from the $N$ antennas will keep near-logarithm increasing with $N$, which corresponds to the simulation in \cite{pizzo2022fourier}. 

According to (\ref{equ_trace_TE}), the noise power in the discrete receiver model can be controlled by
\begin{equation}
	\begin{aligned}
		n_1 = n_0\frac{\sum_{i=1}^m K_E(r_i,r_i)}{\int_0^l K_E(r,r){\rm d}r}.
	\end{aligned}
	\label{equ:n1n0}
\end{equation}

We denote the determinant of matrix ${\bf K} \in \mathbb{C}^{m \times m}$ by 
${\rm det}({\bf K}_{i,j})_{i,j=1}^m$. 
Then we can express the mutual information between the transceivers by 
\ifx\onecol\undefined
\begin{equation}
	\begin{aligned}
	    I_1 &= {\rm log}\left(\frac{{\rm det}({\bf K}_N^{'}+{\bf K}_E^{'})}{{\rm det}({\bf K}_N^{'})}\right) \\&= {\rm log}{\rm det}\left({\mathbbm 1}_{i=j}+\frac{K_E(r_i,r_j)}{n_1/2}\right)_{i,j=1}^{m}.
		\label{ML_discrete_point_antenna}
	\end{aligned}
\end{equation}
\else
\begin{equation}
	\begin{aligned}
	    I_1 &= {\rm log}\left(\frac{{\rm det}({\bf K}_N^{'}+{\bf K}_E^{'})}{{\rm det}({\bf K}_N^{'})}\right) = {\rm log}{\rm det}\left({\mathbbm 1}_{i=j}+\frac{K_E(r_i,r_j)}{n_1/2}\right)_{i,j=1}^{m}.
		\label{ML_discrete_point_antenna}
	\end{aligned}
\end{equation}
\fi

\begin{remark}
	Some annotations about the SNR control: here the SNR on each of the point antennas in the discrete model changes with the density of point antennas. Notice that $\frac{l}{m}\sum_{i=1}^m K_E(r_i,r_i)$ is the approximation of the integral $\int_0^l K_E(r,r){\rm d}r$. When $m$ approximates infinity, $n_1$ will approximate $\frac{mn_0}{2l}$. This phenomenon has several annotations, including the increase of the noise power on each point antenna, the reduction of antenna efficiency, and the corollary of the discretization of EIT continuous models. 
	
	From the perspective of noise power, we can explain it by spatial sampling. For the point antenna arrays, more antennas on a given aperture corresponds to a higher sampling rate in the spatial domain and a wider lowpass filter in the wavenumber domain. Since a wide lowpass filter can receive more noise power from the white noise field, the noise power should increase with the density of the antennas. 

	From the perspective of antenna efficiency, the well-known Hannan's efficiency shows that for both transmitting and receiving antennas, the antenna gain is proportional to $l_xl_y$ for two-dimensional surface antennas~\cite{hannan1964element}. Therefore, for the linear model we considered, the antenna gain will be inversely proportional to the sampling number when the antennas are dense enough. 
	
	\begin{figure*}
		\centering 
		\includegraphics[width=1\linewidth]{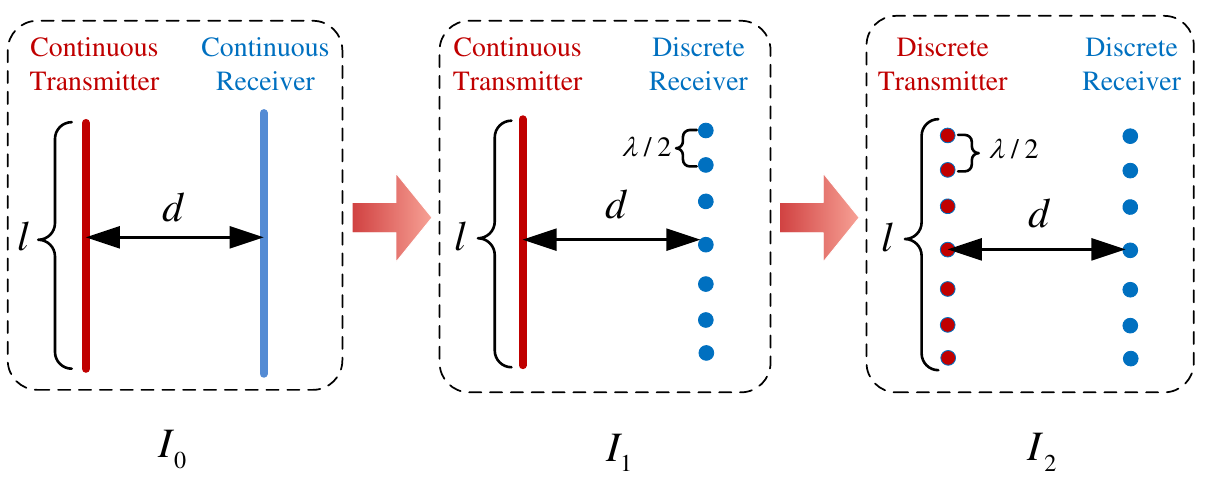} 
		\caption{Comparison between the three models in this section with continuous transceivers and the model with discrete transceivers.} 
		\label{fig_discrete_continuous_merge}
	\end{figure*}

	Besides these two annotations, another perspective is viewing the model of discrete point antennas as the discretization from the EIT continuous model. If we consider $m$ linear continuous antennas instead of point antennas in the destination region. All the antennas are connected head to tail to occupy the $[0, l]$ position in the space and detect the electric field by inner producting it with its eigenmode. This model fulfills the requirement of discretizing the continuous receiver to discrete receiving antennas. The signal received by the $i_{\rm th}$ antenna is 
			$Y_i = \int_{a_i}^{a_{i+1}}Y(r)\phi(r){\rm d}r$, 
	where $[a_i,a_{i+1}]$ is the occupied region of the $i_{\rm th}$ antenna, and $\phi(r)$ is the eigenmode of the antenna. If we assume $\phi(r) \equiv 1$, the correlation matrix of the received electric field can be expressed by 
	\ifx\onecol\undefined
	\begin{equation}
		\begin{aligned}
			({\bf K}_E)_{i,j} &= {\mathbb E}\left[\int_{a_i}^{a_{i+1}}\int_{a_j}^{a_{j+1}} E(r)E^{*}(r'){\rm d}r{\rm d}r'\right] \\&= (a_{i+1}-a_i)(a_{j+1}-a_j)K_E(r_i,r_j),
		\end{aligned}
	\end{equation}
	\else
	\begin{equation}
		\begin{aligned}
			({\bf K}_E)_{i,j} &= {\mathbb E}\left[\int_{a_i}^{a_{i+1}}\int_{a_j}^{a_{j+1}} E(r)E^{*}(r'){\rm d}r{\rm d}r'\right] = (a_{i+1}-a_i)(a_{j+1}-a_j)K_E(r_i,r_j),
		\end{aligned}
	\end{equation}
	\fi
	where $r_i \in [a_i,a_{i+1}]$ and $r_j \in [a_j,a_{j+1}]$ according to the mean value theorem for integrals.
	For the noise field on the destination, we have
	\ifx\onecol\undefined
	\begin{equation}
		\begin{aligned}
			({\bf K}_N)_{i,j} &= {\mathbb E}\left[\int_{a_i}^{a_{i+1}}\int_{a_j}^{a_{j+1}} N(r)N^{*}(r'){\rm d}r{\rm d}r'\right] \\&= \left\{\begin{matrix}
				(a_{i+1}-a_i)\frac{n_0}{2}& i=j\\
				0& i \ne j
			  \end{matrix}\right..
		\end{aligned}
	\end{equation}
	\else
	\begin{equation}
		\begin{aligned}
			({\bf K}_N)_{i,j} &= {\mathbb E}\left[\int_{a_i}^{a_{i+1}}\int_{a_j}^{a_{j+1}} N(r)N^{*}(r'){\rm d}r{\rm d}r'\right] = \left\{\begin{matrix}
				(a_{i+1}-a_i)\frac{n_0}{2}& i=j\\
				0& i \ne j
			  \end{matrix}\right..
		\end{aligned}
	\end{equation}
	\fi
	Therefore, the SNR after the discretization will decrease by $a_{i+1}-a_{i}$, which is the case when the antennas are dense enough.
\end{remark}

After explaining the rationality of the SNR control scheme, we will introduce the following lemma to show the convergence of the noise power on each discrete point antenna, which will be useful for the following proofs.
\begin{lemma}
	When the number of antennas $m$ in a given aperture increases, the noise power on each antenna $n_1/2$ will approach $\frac{mn_0}{2l}$. The difference between them is at most inverse-proportional to $m$. 
\end{lemma}
\begin{IEEEproof}
	From (\ref{equ:n1n0}) and the middle point quadrature rule, we have
	\ifx\onecol\undefined
	\begin{equation}
		\begin{aligned}
			\left|\frac{l}{m}n_1-n_0\right| &= n_0\frac{\left|\int_0^l K_E(r,r){\rm d}r - l/m\sum_{i=1}^m K_E(r_i,r_i)\right|}{\left|\int_0^l K_E(r,r){\rm d}r\right|}
			 \\&\leqslant  \frac{n_0l^3 \left\| K_E^{''}(r,r) \right\|_{L^{\infty}(0,l)}}{24m^2\left|\int_0^l K_E(r,r){\rm d}r\right|},
		\end{aligned}
	\end{equation}
	\else
	\begin{equation}
		\begin{aligned}
			\left|\frac{l}{m}n_1-n_0\right| &= n_0\frac{\left|\int_0^l K_E(r,r){\rm d}r - l/m\sum_{i=1}^m K_E(r_i,r_i)\right|}{\left|\int_0^l K_E(r,r){\rm d}r\right|}
			 \leqslant  \frac{n_0l^3 \left\| K_E^{''}(r,r) \right\|_{L^{\infty}(0,l)}}{24m^2\left|\int_0^l K_E(r,r){\rm d}r\right|},
		\end{aligned}
	\end{equation}
	\fi
	which completes the proof.
\end{IEEEproof}

\subsection{Discrete transceivers}
The models discussed in the above subsections keep the transmitter continuous and only perform discretization on the receiver. 
However, the commonly used model to depict wireless communication is the discrete MIMO model, in which both the transceivers are modeled as discrete point antennas. Therefore, in this section, we will introduce a model which discretizes the transceivers simultaneously. This model can be viewed as the extension of the model with continuous transmitter and discrete receiver. Then, similar to the scheme in the above subsection, we will provide the corresponding SNR control scheme to ensure the fairness of the comparison.

\begin{figure*}
	\centering 
	\includegraphics[width=0.9\linewidth]{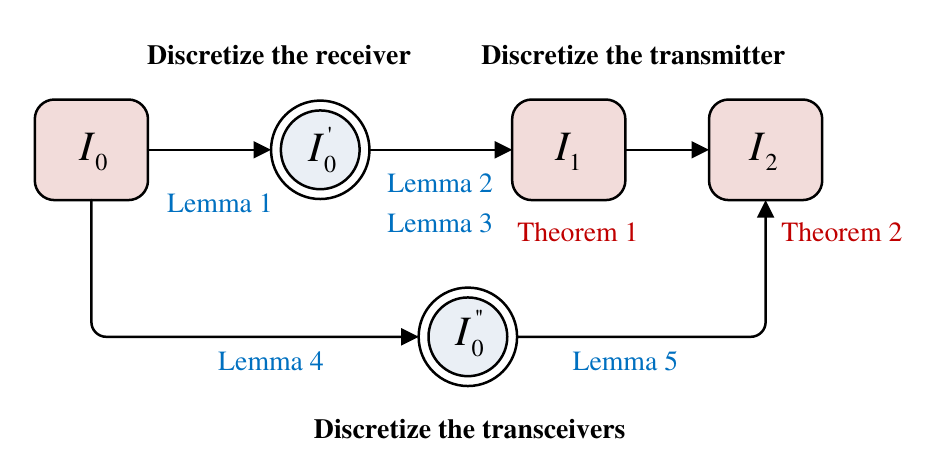} 
	\caption{Flow chart of the proof in this paper.} 
	\label{fig_flow_chart}
\end{figure*}

Specifically, we build a model with $m_1$ point antennas on a length-$l$ segment in the source region and $m_2$ point antennas on a length-$l$ segment in the destination region. Similar to the above subsection, we assume that the $i_{\rm th}$ point antenna is placed at $s_i$ in the source region and $r_i$ in the destination region. The correlation matrix of the signals in the source region is set to be an identity matrix ${\bf K}^{''}_J =P{\bf I}_{m_1}$, which corresponds to the power allocation scheme with no channel state information at the transmitter. The channel gain from the $i_{\rm th}$ antenna in the source region and the $j_{\rm th}$ antenna in the destination region can be expressed by ${\bf H}_{i,j} = G(s_i,r_j)$. The correlation matrix of the received signal is denoted by ${\bf K}^{''}_E  = {\bf H} {\bf K}_{J}^{''} {\bf H}^{\rm H} $. The noise matrix is denoted by ${\bf K}_{N}^{''}=\frac{n_2}{2}{\bf I}_{m_2}$. Similar SNR control on the receiver side is used, which is expressed by 
\begin{equation}
	\begin{aligned}
		n_2 = n_0\frac{\sum_{i=1}^{m_2}\sum_{j=1}^{m_1} G(r_i,s_j)G^{*}(r_i,s_j)}{\int_0^l \int_0^l G(r,s)G^{*}(r,s){\rm d}r{\rm d}s}.
	\end{aligned}
	\label{equ:n2n0}
\end{equation}
The mutual information between the transceivers is expressed as:
\ifx\onecol\undefined
\begin{equation}
	\begin{aligned}
	    I_2 &= {\rm log}\left(\frac{{\rm det}({\bf K}_N^{''}+{\bf K}_E^{''})}{{\rm det}({\bf K}_N^{''})}\right) \\&= {\rm log}{\rm det}\left({\mathbbm 1}_{i=j}+\frac{\sum_{k=1}^{m_1} G(r_i,s_k)G^{*}(r_j,s_k)}{n_2/2}\right)_{i,j=1}^{m_2}.
		\label{ML_double_discrete_point_antenna}
	\end{aligned}
\end{equation}
\else
\begin{equation}
	\begin{aligned}
	    I_2 &= {\rm log}\left(\frac{{\rm det}({\bf K}_N^{''}+{\bf K}_E^{''})}{{\rm det}({\bf K}_N^{''})}\right) = {\rm log}{\rm det}\left({\mathbbm 1}_{i=j}+\frac{\sum_{k=1}^{m_1} G(r_i,s_k)G^{*}(r_j,s_k)}{n_2/2}\right)_{i,j=1}^{m_2}.
		\label{ML_double_discrete_point_antenna}
	\end{aligned}
\end{equation}
\fi
The comparison between the three models built in Section.~\ref{subsec_continuous transceivers}, Section.~\ref{subsec_discrete_receiver} and in this subsection is shown in Fig.~\ref{fig_discrete_continuous_merge}. In the following two sections we will introduce the intermediate quantity $I_0^{'}$ and $I_0^{''}$ to theoretically prove that $I_1$ and $I_2$ converge to $I_0$. The flow chart of the proof is shown in Fig.~\ref{fig_flow_chart}

\section{Performance comparison between discrete and continuous receivers}
\label{sec_discrete_receiver}

In the above section we have proposed the models of continuous and discrete transceivers and derived the corresponding mutual information. Before comparing the performance between CAP-MIMO with continuous transceivers and traditional MIMO with discrete {transceivers}, we first utilize the simplified model with continuous transmitter and discrete {receiver} in this section. Under this simplified model, the transmitter is continuous, which is the same as that in the CAP-MIMO system.
The comparison is based on the convergence analysis of the mutual information when the number of antennas in the discrete receiver increases. Numerical analysis is provided to verify the correctness of the convergence analysis. The discussion about discrete transceivers inspired by the analysis in this part will be in the next section.

\subsection{Convergence analysis of the mutual information}
To compare the mutual information $I_0$ and $I_1$, we introduce an intermediate quantity $I_0^{'} = {\rm log}{\rm det}\left({\bf 1}+\frac{mT_E}{ln_1/2}\right)$. We can bound $|I_0-I_1|$ by $|I_0-I_0^{'}|+|I_1-I_0^{'}|$. According to \cite{bornemann2010numerical}, $I_1$ can be viewed as the approximation of $I_0^{'}$ using a numerical integral scheme. In our discussion the point antennas in the destination region are evenly spaced, which means that $a_i = (i-1)l/m$ and $r_i = (i-0.5)l/m$. 

To bound $|I_1-I_0^{'}|$, we introduce the following lemma 
from \cite{bornemann2010numerical}:
\begin{lemma}
	We define $d(z) := {\rm det}({\bf 1}+zT)$ and $d_Q(z):={\rm det}\left({\mathbbm 1}_{i=j}+w_j z K(r_i,r_j)\right)_{i,j=1}^{m}$, where $K$ is the kernel of the integral operator $T: = \phi(x) \rightarrow \int_{[a,b]}K(x,y)\phi(y){\rm d}y$.  
	The difference between $d(z)$ and $d_Q(z)$ is 
	\ifx\onecol\undefined
	\begin{equation}
		\begin{aligned}
	d(z)-d_Q(z) &= \sum_{n=1}^{\infty}\frac{z^n}{n!}\Bigg( Q^n_m(K_n)\\&-\int_{[a,b]^n}K_n(x_1,\cdots,x_n){\rm d}x_1\cdots{\rm d}x_n \Bigg),
\end{aligned}
\label{equ_lemma2}
\end{equation}
\else
\begin{equation}
	\begin{aligned}
d(z)-d_Q(z) = \sum_{n=1}^{\infty}\frac{z^n}{n!}\Bigg( Q^n_m(K_n)-\int_{[a,b]^n}K_n(x_1,\cdots,x_n){\rm d}x_1\cdots{\rm d}x_n \Bigg),
\end{aligned}
\label{equ_lemma2}
\end{equation}
\fi
where $K_n(x_1,\cdots,x_n) = {\rm det}\left( K(x_i,x_j) \right)_{i,j=1}^n$, and $Q_m^n(f) = \sum_{j_1=1,\cdots,j_n=1}^m \prod_{i=1}^{n}w_{j_i}f(r_{j_1},\cdots,r_{j_n})$.
\end{lemma}

{\bf Lemma 2} provides a method to compare the difference between a Fredholm determinant of operator and a classical determinant of matrix. In our model, the operator $T$ corresponds to the integral operator $T_E$, $z$ equals $\frac{2m}{ln_1}$,and $w_j = l/m$ according to the equally spaced antennas.  

Notice that $Q_m^n(f)$ is the numerical approximation of the integral $\int_{[a,b]^n}K_n(x_1,\cdots,x_n){\rm d}x_1\cdots{\rm d}x_n$, we need to use numerical integral theory to estimate the approximation error. For the model with equally spaced antennas, this expression corresponds to a multivariate $m$-point composite midpoint quadrature rule. 

	For the error bound of a $m$-point composite midpoint quadrature \cite{davis2007methods}, we have
	\begin{equation}
		\begin{aligned}
			\left| Q_m(f)-\int_0^l f(x) {\rm d}x \right| \leqslant \frac{l^3}{24m^2}\| f^{''} \|_{L^{\infty}(0,l)}
		\end{aligned}
	\end{equation}
	
	According to \cite{haber1970numerical}, the numerical approximation error for multiple integrals in a $n$-dimensional unit cube can be bounded by
	\begin{equation}
		\begin{aligned}
			\left| \int_{G_n}f - \left(\prod_{i=1}^n\right)Q_i(f;x_i)  \right| \leqslant E_1+\sum_{i=2}^n \prod_{j=1}^{i-1} W_j E_i,
		\end{aligned}
	\end{equation}
	where 
		\begin{equation}
		\begin{aligned}
	Q_i(f;x_i):= \sum_{j_i=1}^m w_{j_i}f(x_1,\cdots,x_i=r_{j_i},\cdots,x_n), 
			\end{aligned}
\end{equation}
$W_i = \sum_{j_i} |w_{j_i}|$ ,
	and $E_i \geqslant \left| Q_i(f;x_i) - \int_0^1 f(x_1,\cdots,x_n){\rm d}x_i  \right|$. According to the models in this paper, we have $w_{i,j} = l/m$ and $W_i = l$. 
	By simple variation of the integral band, we can bound the approximation error of the multi-dimensional numerical integral quadrature rule by 
	\ifx\onecol\undefined
	\begin{equation}
		\begin{aligned}
			&\left| Q_m^n(K_n)-\int_{[0,l]^n} K_n(x_1,\cdots,x_n) {\rm d}x_1 \cdots {\rm d}x_n \right| \\&\leqslant l^{n-1} \sum_{i=1}^n E_i,
		\end{aligned}
	\end{equation}
	\else
	\begin{equation}
		\begin{aligned}
			\left| Q_m^n(K_n)-\int_{[0,l]^n} K_n(x_1,\cdots,x_n) {\rm d}x_1 \cdots {\rm d}x_n \right| \leqslant l^{n-1} \sum_{i=1}^n E_i,
		\end{aligned}
	\end{equation}
	\fi
	where 
	\begin{equation}
		\begin{aligned}
			E_i = \left| Q_i(K_n;x_i) - \int_0^l K_n(x_1,\cdots,x_n){\rm d}x_i \right|.
		\end{aligned}
	\end{equation}
	Therefore, we have
	\ifx\onecol\undefined
	\begin{equation}
		\begin{aligned}
			&\left| Q_m^n(K_n)-\int_{[0,l]^n} K_n(x_1,\cdots,x_n) {\rm d}x_1 \cdots {\rm d}x_n \right| \\&\leqslant  \frac{nl^{n+2}}{24 m^2}\left| K_n \right|_2
		\end{aligned}
		\label{equ_multivariate_quadrature}
	\end{equation}
	\else
	\begin{equation}
		\begin{aligned}
			\left| Q_m^n(K_n)-\int_{[0,l]^n} K_n(x_1,\cdots,x_n) {\rm d}x_1 \cdots {\rm d}x_n \right| \leqslant  \frac{nl^{n+2}}{24 m^2}\left| K_n \right|_2
		\end{aligned}
		\label{equ_multivariate_quadrature}
	\end{equation}
	\fi
	where $|K_n|_2 = \underset{i}{{\rm max}} \| \frac{\partial^2 K_n}{\partial x_{i}^2} \|_{L^\infty((0,l)^n)}$.

Similar to \cite[{\bf Lemma A.4}]{bornemann2010numerical}, we can bound $|K_n|_k$ by using the Hadamard's inequality, which leads to
	\begin{equation}
		\begin{aligned}
	\left| K_n  \right|_k \leqslant 2^k n^{n/2} \left( \underset{i+j\leqslant k}{{\rm max}}\left\| \frac{\partial_{x}^i \partial_{y}^j K(x,y)}{\partial x^i\partial y^j} \right\|_{L^{\infty}((0,l)^2)} \right)^n.
\end{aligned}
\label{equ_Kn}
\end{equation}

Next we will show that $\left| \frac{\partial_{x}^i \partial_{y}^j K(x,y)}{\partial x^i\partial y^j} \right|$ is upper-bounded. Since we have $K(x,y) = \int_0^l G(x,s)G^{*}(y,s) {\rm d}s$, we will first analyze the property of $G(x,s)$. We decompose $G(x,s)$ as $G_1(x,s)+{\rm j}G_2(x,s)$, where $G_1, G_2\in C^{\infty}([0,l]^2)$. The smoothness of $G_1, G_2$ in their domains is trivial since they are compositions of polynomial functions, trigonometric functions and square root functions. 
Consider the integral kernel $K(x,y)$ expressed in terms of $G_1, G_2$, i.e., 
\ifx\onecol\undefined
\begin{equation}
	\begin{aligned}
K(x,y) &= \int_0^l \big( G_1(x,s)G_1(y,s) +G_2(x,s)G_2(y,s)\big){\rm d}s \\&+ {\rm j}\int_0^l \big(G_1(y,s)G_2(x,s)-G_1(x,s)G_2(y,s)\big){\rm d}s.
	\end{aligned}
\end{equation}
\else
\begin{equation}
	\begin{aligned}
K(x,y) = \int_0^l \big( G_1(x,s)G_1(y,s) +G_2(x,s)G_2(y,s)\big){\rm d}s + {\rm j}\int_0^l \big(G_1(y,s)G_2(x,s)-G_1(x,s)G_2(y,s)\big){\rm d}s.
	\end{aligned}
\end{equation}
\fi
Since $G_1(x,s)$ and $G_2(y,s)$ are smooth in $[0,l]^2$, we can conclude that $f_1(x,y) = G_1(x,s)G_1(y,s) +G_2(x,s)G_2(y,s)$ and $f_2(x,y) = G_1(y,s)G_2(x,s)-G_1(x,s)G_2(y,s)$ are smooth in the same domain. 
Since compactly supported smooth functions attain their maximum or minimum values, the partial derivatives of $K(\cdot, \cdot)$ are upper-bounded for any order $i,j$, i.e., 
\begin{equation}
	\left| \frac{\partial^{i+j} K(x,y)}{\partial x^i\partial y^j} \right| < \infty, \quad\forall i,j.
\end{equation}

Therefore, by substituting (\ref{equ_multivariate_quadrature}) and (\ref{equ_Kn}) into {\bf Lemma 2}, we can bound the difference between the mutual information $I_0^{'}$ and $I_1$ by the following lemma:
\begin{lemma}
	The mutual information $I_1$  converges to the mutual information $I_0^{'}$. The difference $\left| I_1- I_0^{'}\right|$ is at most inverse-proportional to $m^2$.
\end{lemma}
\begin{IEEEproof}
	From (\ref{equ:R_E_noCSI}) we know that for the operator $T_E$, the kernel function can be expressed by $K(x,y) = \int_0^Lg(x,s)g^*(y,s){\rm d}s$. From (\ref{equ_Kn}) we have
	\begin{equation}
		\begin{aligned}
			\left| K_n \right|_2 \leqslant 4n^{n/2}A^n,
		\end{aligned}
	\end{equation}	
	where $A = {\rm max}\left\| \frac{\partial^{i+j} K(x,y)}{\partial x^i\partial y^j} \right\|_{L^{\infty}((0,l)^2)}$ is a constant. Therefore we have
	\ifx\onecol\undefined
	\begin{equation}
		\begin{aligned}
	\left|d(z)-d_Q(z)\right| & \leqslant \sum_{n=1}^{\infty}\frac{z^n}{n!}\frac{nl^{n+2}}{24m^2}\underset{i}{{\rm max}} \left\| \frac{\partial^2 K_n}{\partial x^2} \right\|_{L^\infty((0,l)^n)}
	 \\&\leqslant \sum_{n=1}^{\infty}\frac{z^n}{n!}\frac{nl^{n+2}}{6m^2}n^{n/2}A^n.
\end{aligned}
\end{equation}
\else
\begin{equation}
	\begin{aligned}
\left|d(z)-d_Q(z)\right|  \leqslant \sum_{n=1}^{\infty}\frac{z^n}{n!}\frac{nl^{n+2}}{24m^2}\underset{i}{{\rm max}} \left\| \frac{\partial^2 K_n}{\partial x^2} \right\|_{L^\infty((0,l)^n)}
 \leqslant \sum_{n=1}^{\infty}\frac{z^n}{n!}\frac{nl^{n+2}}{6m^2}n^{n/2}A^n.
\end{aligned}
\end{equation}
\fi
According to the Stirling's approximation, we have $n! \geqslant n^ne^{-n}\sqrt{2\pi n}$, which leads to
\begin{equation}
	\begin{aligned}
\left|d(z)-d_Q(z)\right| \leqslant \frac{l^2}{6m^2}\sum_{n=1}^{\infty}\sqrt{\frac{n}{2\pi}}\frac{(Aezl)^n}{n^{n/2}}.
\end{aligned}
\label{equ:difference between det}
\end{equation}
	Since it is obvious that $\sum_{n=1}^{\infty}\sqrt{\frac{n}{2\pi}}\frac{(Aezl)^n}{n^{n/2}}$ is convergent, the difference between $d(z)$ and $d_Q(z)$ is proportional to $m^{-2}$. For the difference between mutual information $I_1$ and $I_0^{'}$, we have
	\begin{equation}
		|I_1-I_0^{'}| \leqslant \frac{|d(z)-d_Q(z)|}{{\rm min}(d(z),d_Q(z))} < \frac{l^2}{6m^2}\sum_{n=1}^{\infty}\sqrt{\frac{n}{2\pi}}\frac{(Aezl)^n}{n^{n/2}},
		\label{equ:I1I2^}
	\end{equation}
	where $z = \frac{m}{ln_1/2}$. From {\bf Lemma 1} we know that $\frac{l}{m}n_1\geqslant n_0 - \frac{n_0l^3 \left\| K_E^{''}(r,r) \right\|_{L^{\infty}(0,l)}}{24m^2\left|\int_0^l K_E(r,r){\rm d}r\right|}$. Therefore, $z$ is upperbounded, which completes the proof of {\bf Lemma 3}.
\end{IEEEproof}

According to {\bf Lemma 1} and {\bf Lemma 3}, we have {\bf Theorem 1}, which bounds the difference between $I_0$ and $I_1$.

\begin{theorem}
	The mutual information $I_1$ that can be obtained from the discrete receiver converges to the mutual information $I_0$ that can be obtained from the continuous receiver when the number of points increases. The difference $|I_1-I_0|$ is at most inverse-proportional to the square of the sampling number $m$.
\end{theorem}

\begin{IEEEproof}
	Since the Fredholm determinant $f(z) = {\rm det}({\bf 1}+zT_E)$ is an analytic function, we have 
	\ifx\onecol\undefined
	\begin{equation}
		\begin{aligned}
			&~~~\left| {\rm det}({\bf 1}+zT_E)-{\rm det}({\bf 1}+z_1T_E)\right| \\&= |z-z_1| \left| \frac{\partial {\rm det}({\bf 1}+xT_E)}{\partial x}\right|_{x \in [{\rm min}(z,z_1),{\rm max}(z,z_1)]}.
		\end{aligned}
		\label{equ:z_variate}
	\end{equation}
	\else
	\begin{equation}
		\begin{aligned}
			\left| {\rm det}({\bf 1}+zT_E)-{\rm det}({\bf 1}+z_1T_E)\right| = |z-z_1| \left| \frac{\partial {\rm det}({\bf 1}+xT_E)}{\partial x}\right|_{x \in [{\rm min}(z,z_1),{\rm max}(z,z_1)]}.
		\end{aligned}
		\label{equ:z_variate}
	\end{equation}
	\fi
	In our assumption $z = \frac{2}{n_0}$ and $z_1 = \frac{2m}{ln_1}$. The analycity of ${\rm det}({\bf 1}+zT_E)$ implies that $\frac{\partial {\rm det}({\bf 1}+xT_E)}{\partial x}$ is also an analytic function and is bounded on the interval $[{\rm min}(z,z_1),{\rm max}(z,z_1)]$. We denote $M = \underset{x}{\rm max}\left| \frac{\partial {\rm det}({\bf 1}+xT_E)}{\partial x} \right|$, where ${x \in [{\rm min}(z,z_1),{\rm max}(z,z_1)]}$.
	From {\bf Lemma 1} we have 	
	\begin{equation}
		\begin{aligned}
			\left|\frac{l}{m}n_1-n_0\right| \leqslant  \frac{n_0l^3 \left\| K_E^{''}(r,r) \right\|_{L^{\infty}(0,l)}}{24m^2\int_0^l K_E(r,r){\rm d}r}.
		\end{aligned}
	\end{equation}
	Since $n_1/m \rightarrow n_0/l$ when $m$ approximates infinity, we denote the minimum value of $n_1/m$ by $c$.
	Therefore, $\left| {\rm det}({\bf 1}+\frac{2}{n_0}T_E)-{\rm det}({\bf 1}+\frac{2m}{ln_1}T_E)\right| $ can be bounded by
	\ifx\onecol\undefined
	\begin{equation}
		\begin{aligned}
			&~~~\left| {\rm det}({\bf 1}+\frac{2}{n_0}T_E)-{\rm det}({\bf 1}+\frac{2m}{ln_1}T_E)\right| \\&\leqslant \frac{M l^2 \left\| K_E^{''}(r,r) \right\|_{L^{\infty}(0,l)}}{12m^2 c \int_0^l K_E(r,r){\rm d}r}.
		\end{aligned}
	\end{equation}
	\else
	\begin{equation}
		\begin{aligned}
			\left| {\rm det}({\bf 1}+\frac{2}{n_0}T_E)-{\rm det}({\bf 1}+\frac{2m}{ln_1}T_E)\right| \leqslant \frac{M l^2 \left\| K_E^{''}(r,r) \right\|_{L^{\infty}(0,l)}}{12m^2 c \int_0^l K_E(r,r){\rm d}r}.
		\end{aligned}
	\end{equation}
	\fi
	Similar to the derivation of (\ref{equ:I1I2^}), we know that when $m$ increases, $I_0^{'}$ will converge to $I_0$. The difference is at most inverse-proportional to $m^2$. From {\bf Lemma 3} we know that $\left| I_1-I_0^{'} \right|$ is at most inverse-proportional to $m^2$. Since $\left| I_0-I_1 \right| \leqslant \left| I_0-I_0^{'} \right|+\left| I_1-I_0^{'} \right|$, {\bf Theorem 1} is proved.
\end{IEEEproof}

\begin{remark}
	Self-consistency of the model: {\bf Theorem 1} shows that the SNR control scheme between the discrete and continuous models is appropriate, since the limit of the mutual information of the discrete model is proved to be that of the continuous model. That is to say, our proposed model is self-consistent. Therefore, we can use the proposed model to compare the mutual information from the discrete and continuous receivers. Our analysis is based on $R_E(r,r') = P\int_0^l G(r,s)G^*(r',s){\rm d}s$ which corresponds to the scenario when no CSI can be obtained at the transmitter but not limited to this scenario. It can be easily extended to other shapes of autocorrelation functions after power allocation at the transmitter, as long as the analycity of $R_E(r,r')$ is guaranteed. 
\end{remark}

\begin{remark}
	Extension of the convergence analysis: In {\bf Remark 1} we have shown that the mutual information analysis can be naturally extended to other scenarios besides the linear transceivers considered in our paper. Based on the mutual information analysis, we can also extend the convergence analysis we have made in this subsection. For example, we can show that the mutual information between discretized two-dimensional transceivers converges to the mutual information between continuous two-dimensional transceivers. Notice that the convergence analysis is mainly based on the series expansion of determinant in {\bf Lemma 2}, where $T$ is the integral operator of one variable and $K$ is in the matrix form. If we extend $T$ to the integral operator with several variables and $K$ to a multi-dimensional tensor, we can find an extension of {\bf Lemma 2}, and the other parts of the convergence proof will have no essential difference with what we have done in this section. We show in {\bf Appendix A} how this extension works and omit the other parts of the convergence proof.
\end{remark}

\subsection{Numerical analysis about the mutual information}
As proven in the above subsection, the mutual information between the continuous transmitter and discrete receiver converges to the mutual information between continuous transceivers. Therefore, the model of the discrete receiver can be viewed as the discretization of the continuous receiver. In this subsection, we will use numerical analysis to show the correctness of the theoretical results. Moreover, we will show the near-optimality of the discrete receiver with half-wavelength sampling.

\ifx\onecol\undefined
\begin{figure}
	\centering 
	\includegraphics[height=7cm, width=9cm]{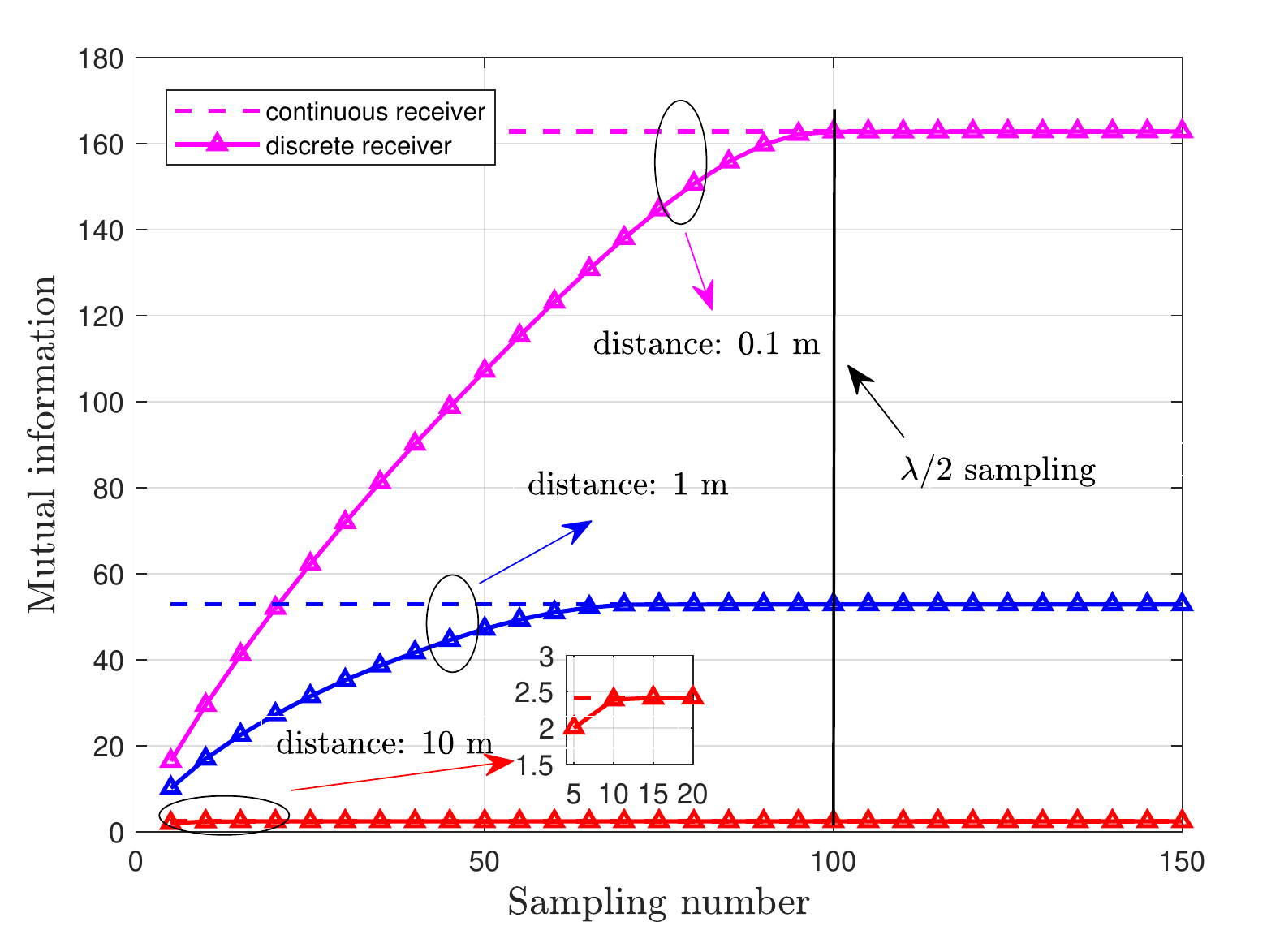} 
	\caption{The mutual information as a function of the sampling number. The transmitter is kept continuous and the receiver is discretized.} 
	\label{fig_discrete_receiver}
\end{figure}
\else
\begin{figure*}
	\centering 
	\includegraphics[height=10cm, width=13cm]{figs1/discrete_receiver.pdf} 
	\caption{The mutual information as a function of the sampling number. The transmitter is kept continuous and the receiver is discretized.} 
	\label{fig_discrete_receiver}
\end{figure*}
\fi

We set the length $l$ of the transceivers to 2\,$\rm m$. The transmitter is kept continuous, while the receiver is discretized to $m$ point antennas. The wavelength of the electromagnetic field is fixed to 0.04\,$\rm m$, which corresponds to the frequency of 7.5\,$\rm GHz$.
The distance between the transceivers varies from 10\,$\rm m$ to 0.1\,$\rm m$. The simulation results are shown in Fig. \ref{fig_discrete_receiver}. From the simulation, we can observe the convergence of the mutual information between the continuous transmitter and the discrete receiver, which verifies the theoretical analysis. For the three distances between transceivers, the half-wavelength sampling almost achieves the supremum mutual information between continuous transceivers. Therefore, half-wavelength sampling of the receiver is suboptimal. Moreover, when the distance between transceivers decreases, we can observe that the mutual information converges slower. When the distance equals 0.1\,$\rm m$, the half wavelength sampling is at the critical state of convergence. If the distance is less than 0.1\,$\rm m$, a performance gap between the model with the continuous receiver and that with the discrete receiver may be observed. This performance gap has theoretical meaning but may not be useful because the distance will be comparable to the wavelength in this scenario, where the evanescent wave components will hold a dominant position.  
\ifx\onecol\undefined
\begin{figure*}
	\begin{equation}
		\begin{aligned}
			&~~\left|  n_0 - \frac{l^2}{m_2 m_1}n_2  \right|  = \frac{n_0}{\int_0^lK(r,r){\rm d}r}\left|  \int_0^l\int_0^l g(r,r,s){\rm d}s{\rm d}r - \frac{l^2}{m_1 m_2}\sum_{i=1,j=1}^{m_2,m_1}g(r_i,r_i,s_j) \right| 
			\\& \leqslant \frac{n_0 l^4}{24m_2^2 \int_0^l K(r,r){\rm d}r} \left\| \frac{\partial^2 g(r,r,s)}{\partial r^2} \right\|_{L^{\infty}((0,l)^2)} + \frac{n_0 l^4}{24m_1^2 \int_0^l K(r,r){\rm d}r}\left\| \frac{\partial^2 g(r,r,s)}{\partial s^2} \right\|_{L^{\infty}((0,l)^2)}  
			\\&\leqslant \frac{n_0 l^4}{24 ({\rm min}(m_1,m_2))^2 \int_0^l K(r,r){\rm d}r}\left( \left\| \frac{\partial^2 g(r,r,s)}{\partial r^2} \right\|_{L^{\infty}((0,l)^2)} +\left\| \frac{\partial^2 g(r,r,s)}{\partial s^2} \right\|_{L^{\infty}((0,l)^2)}  \right), 
		\end{aligned}
		\label{equ_n0_m1m2_n2}
	\end{equation}
	{\noindent} \rule[-10pt]{18cm}{0.05em}
\end{figure*}
\else
\fi

\section{Comparison between continuous and discrete transceivers}
\label{sec_discrete_transceiver}
In the above section we have compared the mutual information between the models with continuous and discrete receivers. For both models the transmitter is kept continuous, which simplifies the analyzing procedure. Inspired by the analysis in the above section, in this section we will compare the mutual information between continuous transceivers and that between discrete transceivers. Numerical analysis is then provided to show the near-optimality of the half-wavelength sampling scheme.
\subsection{Convergence analysis of the mutual information}
\sloppy The analysis in this section focuses on the difference between $I_0$ and $I_2$. It is an extension of the convergence analysis in the above section. We define $I_0^{''} = {\rm logdet}\left( {\bf 1} + \frac{m_1 m_2 T_E}{l^2n_2/2} \right)$ as an intermediate variable similar to $I_0^{'}$.
First we will discuss the convergence of $|I_0-I_0^{''}|$ in the following lemma:
\begin{lemma} 
	The mutual information $I_0^{''}$  converges to the mutual information $I_0$. The difference $\left| I_0- I_0^{''}\right|$ is at most inverse-proportional to $({\rm min}(m_1,m_2))^2$.
\end{lemma}
\begin{IEEEproof}
	From the SNR control scheme of discrete transceivers (\ref{equ:n2n0}) and the multivariate $m$-point composite midpoint quadrature rule, we have
	\ifx\onecol\undefined
	(\ref{equ_n0_m1m2_n2})
\else
\begin{equation}
	\begin{aligned}
		&~~\left|  n_0 - \frac{l^2}{m_2 m_1}n_2  \right|  = \frac{n_0}{\int_0^lK(r,r){\rm d}r}\left|  \int_0^l\int_0^l g(r,r,s){\rm d}s{\rm d}r - \frac{l^2}{m_1 m_2}\sum_{i=1,j=1}^{m_2,m_1}g(r_i,r_i,s_j) \right| 
		\\& \leqslant \frac{n_0 l^4}{24m_2^2 \int_0^l K(r,r){\rm d}r} \left\| \frac{\partial^2 g(r,r,s)}{\partial r^2} \right\|_{L^{\infty}((0,l)^2)} + \frac{n_0 l^4}{24m_1^2 \int_0^l K(r,r){\rm d}r}\left\| \frac{\partial^2 g(r,r,s)}{\partial s^2} \right\|_{L^{\infty}((0,l)^2)}  
		\\&\leqslant \frac{n_0 l^4}{24 ({\rm min}(m_1,m_2))^2 \int_0^l K(r,r){\rm d}r}\left( \left\| \frac{\partial^2 g(r,r,s)}{\partial r^2} \right\|_{L^{\infty}((0,l)^2)} +\left\| \frac{\partial^2 g(r,r,s)}{\partial s^2} \right\|_{L^{\infty}((0,l)^2)}  \right), 
	\end{aligned}
	\label{equ_n0_m1m2_n2}
\end{equation}
\fi
where $g(x,y,z) := G(x,z)G^{*}(y,z)$, $r_i = (i-0.5)l/{m_2}$, and $s_j = (j-0.5)l/{m_1}$.
It is obvious that $n_2/(m_1m_2)$ converges to $n_0/l^2$ when $m_1\to\infty$ and $m_2\to\infty$. We denote the minimum value of $n_2/(m_1 m_2)$ by $c$. Then, according to (\ref{equ:z_variate}), we know that $\left| {\rm det}({\bf 1}+\frac{2}{n_0}T_E)-{\rm det}({\bf 1}+\frac{2m_1m_2}{l^2n_2}T_E)\right| $ converges to 0 when $m_1\to\infty$ and $m_2\to\infty$. Therefore, $|I_0-I_0^{''}|$ converges to 0, and the difference is at most inversely proportional to $({\rm min}(m_1,m_2))^2$.
\end{IEEEproof}

Then we will discuss the convergence of $|I_2-I_0^{''}|$ in the following lemma:
\begin{lemma}
	The difference $\left| I_2- I_0^{''}\right|$ approaches 0 when $m$ approaches infinity. Moreover, it is at most inverse-proportional to $({\rm min}(m_1,m_2))^2$. 
\end{lemma}
\begin{IEEEproof}
 We denote the Fredholm determinant and its discretization by $d(z) = {\rm det}({\bf 1}+zT)$ and $d_V(z)={\rm det}\left({\mathbbm 1}_{i=j}+w'_j z \sum_{k=1}^{m_1} w_k G(r_i,s_k)G^{*}(r_j,s_k)  \right)_{i,j=1}^{m_2}$, where $K(x_i,x_j) = \int_0^lG(x_i,s)G^{*}(x_j,s){\rm d}s $ is the kernel of the operator $T$. To bound the difference between $d(z)$ and $d_V(z)$,  we define 
 $g_n(x_1,\cdots,x_n,s_1,\cdots,s_n)$ as
 \ifx\onecol\undefined
 \begin{equation}
	\begin{aligned}
		&~~~g_n(x_1,\cdots,x_n,s_1,\cdots,s_n) \\&= {\rm det}
		\begin{bmatrix}
			g(x_1,x_1,s_1)&\cdots&g(x_1,x_n,s_1)\\
			\cdots & g(x_i,x_j,s_i)&\cdots\\
			g(x_n,x_1,s_n)&\cdots&g(x_n,x_n,s_n)
		\end{bmatrix}.
	\end{aligned}
 \end{equation}
 \else
 \begin{equation}
	\begin{aligned}
		g_n(x_1,\cdots,x_n,s_1,\cdots,s_n) = {\rm det}
		\begin{bmatrix}
			g(x_1,x_1,s_1)&\cdots&g(x_1,x_n,s_1)\\
			\cdots & g(x_i,x_j,s_i)&\cdots\\
			g(x_n,x_1,s_n)&\cdots&g(x_n,x_n,s_n)
		\end{bmatrix}.
	\end{aligned}
 \end{equation}
 \fi

 From the definition of $g(x,y,z)$, we know that $\int_0^l g(x_i,x_j,s_i){\rm d}s_i = K(x_i,x_j) $.
 According to the property of determinants that ${\rm det}(a_{i,j})_{i,j=1}^m = \sum_{\pi \in \sigma_m} (-1)^{\pi} a_{1,\pi(1)}\cdots a_{m,\pi(m)}$, where $\sigma_m$ is the family of all permutations on ${1,\cdots,m}$ and $(-1)^{\pi}$ is the sign of the permutation $\pi$,
 we can find that 
 \ifx\onecol\undefined
 \begin{equation}
	\begin{aligned}
 K_n(x_1,\cdots,x_n) =& \int_{[0,l]^n}g_n(x_1,\cdots,x_n,s_1,\cdots,s_n)\\&{\rm d}s_1\cdots{\rm d}s_n.
\end{aligned}
\end{equation}
\else
\begin{equation}
	\begin{aligned}
 K_n(x_1,\cdots,x_n) = \int_{[a,b]^n}g_n(x_1,\cdots,x_n,s_1,\cdots,s_n){\rm d}s_1\cdots{\rm d}s_n.
\end{aligned}
\end{equation}
\fi

If we define $C_{m_1}^n(g_n)$ by 
\ifx\onecol\undefined
(\ref{equ_Cmn})
\begin{figure*}
\begin{equation}
	\begin{aligned}
		C_{m_1}^n(g_n) = {\rm det}
		\begin{bmatrix}
			\sum_{\alpha_1=1}^{m_1} w_{\alpha_1} g(x_1,x_1,s_{1,{\alpha_1}})&\cdots&\sum_{\alpha_1=1}^{m_1} w_{\alpha_1} g(x_n,x_n,s_{1,{\alpha_1}})\\
			\cdots & \sum_{\alpha_i=1}^{m_1} w_{\alpha_i} g(x_i,x_j,s_{i,{\alpha_i}})&\cdots\\
			\sum_{\alpha_n=1}^{m_1} w_{\alpha_n} g(x_n,x_1,s_{n,{\alpha_n}})&\cdots&\sum_{\alpha_n=1}^{m_1} w_{\alpha_n} g(x_n,x_n,s_{n,{\alpha_n}})
		\end{bmatrix},
	\end{aligned}
	\label{equ_Cmn}
 \end{equation}
 {\noindent} \rule[-10pt]{18cm}{0.05em}
\end{figure*}
 \else
 \begin{equation}
	\begin{aligned}
		C_{m_1}^n(g_n) = {\rm det}
		\begin{bmatrix}
			\sum_{\alpha_1=1}^{m_1} w_{\alpha_1} g(x_1,x_1,s_{1,{\alpha_1}})&\cdots&\sum_{\alpha_1=1}^{m_1} w_{\alpha_1} g(x_n,x_n,s_{1,{\alpha_1}})\\
			\cdots & \sum_{\alpha_i=1}^{m_1} w_{\alpha_i} g(x_i,x_j,s_{i,{\alpha_i}})&\cdots\\
			\sum_{\alpha_n=1}^{m_1} w_{\alpha_n} g(x_n,x_1,s_{n,{\alpha_n}})&\cdots&\sum_{\alpha_n=1}^{m_1} w_{\alpha_n} g(x_n,x_n,s_{n,{\alpha_n}})
		\end{bmatrix},
	\end{aligned}
	\label{equ_Cmn}
 \end{equation}
 \fi
 we can obtain 
 \ifx\onecol\undefined
 (\ref{equ_Cmn_expression}).
\begin{figure*}
 \begin{equation}
	\begin{aligned}
		C_{m_1}^n(g_n)  &= \sum_{\pi \in \sigma_n} (-1)^{\pi} \left(\sum_{\alpha_1=1}^{m_1} w_{\alpha_1} g(x_1,x_{\pi(1)},s_{1,{\alpha_1}})\right)\cdots \left(\sum_{\alpha_n=1}^{m_1} w_{\alpha_n} g(x_n,x_{\pi(n)},s_{n,{\alpha_n}})\right)
		\\& = \sum_{\pi \in \sigma_n} \left( (-1)^{\pi}  \sum_{\alpha_1, \cdots, \alpha_n=1}^{m_1} \prod_{i=1}^n w_{\alpha_i}  g(x_i,x_{\pi(i)},s_{i,\alpha_i}) \right)
		\\& = \sum_{\alpha_1, \cdots, \alpha_n=1}^{m_1} \left( \left(\prod_{i=1}^n w_{\alpha_i}\right) \sum_{\pi \in \sigma_n}(-1)^{\pi} \prod_{i=1}^n g(x_i,x_{\pi(i)},s_{i,\alpha_i})\right)
		\\& = \sum_{\alpha_1, \cdots, \alpha_n=1}^{m_1} \left(\left(\prod_{i=1}^n w_{\alpha_i}\right) g_n(x_1,\cdots,x_n,s_{1,\alpha_1},\cdots,s_{n,\alpha_n})\right).
	\end{aligned}
	\label{equ_Cmn_expression}
 \end{equation}
 {\noindent} \rule[-10pt]{18cm}{0.05em}
\end{figure*}
 \else
 \begin{equation}
	\begin{aligned}
		C_{m_1}^n(g_n)  &= \sum_{\pi \in \sigma_n} (-1)^{\pi} \left(\sum_{\alpha_1=1}^{m_1} w_{\alpha_1} g(x_1,x_{\pi(1)},s_{1,{\alpha_1}})\right)\cdots \left(\sum_{\alpha_n=1}^{m_1} w_{\alpha_n} g(x_n,x_{\pi(n)},s_{n,{\alpha_n}})\right)
		\\& = \sum_{\pi \in \sigma_n} \left( (-1)^{\pi}  \sum_{\alpha_1, \cdots, \alpha_n=1}^{m_1} \prod_{i=1}^n w_{\alpha_i}  g(x_i,x_{\pi(i)},s_{i,\alpha_i}) \right)
		\\& = \sum_{\alpha_1, \cdots, \alpha_n=1}^{m_1} \left( \left(\prod_{i=1}^n w_{\alpha_i}\right) \sum_{\pi \in \sigma_n}(-1)^{\pi} \prod_{i=1}^n g(x_i,x_{\pi(i)},s_{i,\alpha_i})\right)
		\\& = \sum_{\alpha_1, \cdots, \alpha_n=1}^{m_1} \left(\left(\prod_{i=1}^n w_{\alpha_i}\right) g_n(x_1,\cdots,x_n,s_{1,\alpha_1},\cdots,s_{n,\alpha_n})\right).
	\end{aligned}
	\label{equ_Cmn_expression}
 \end{equation}
 \fi
Here $w_{\alpha_i}$ corresponds to the distance between antennas in the source region and $s$ corresponds to the location of the antennas in the source region. When further considering the discretization of the receiver as in (\ref{equ_lemma2}), we should set $x_1^n$ to the location of the antennas in the destination region, and add additional weight $w$ which equals the distance between antennas in the destination region. Similar to the definition of $Q_m^n$ in (\ref{equ_lemma2}), we define $V_m^n(g_n)$ by $V_m^n(g_n) = \sum_{j_1,\cdots,j_n=1}^{m_2} \prod_i w'_{j_i} C_{m_1}^n(g_n(r_{j_1},\cdots,r_{j_n},s_{1,\alpha_1},\cdots,s_{n,\alpha_n}))$. When $s_{j,\alpha_i} = s_{\alpha_i}$, we have 
\ifx\onecol\undefined
 \begin{equation}
	\begin{aligned}
 V_m^n(g_n) & = \sum_{j_1,\cdots,j_n=1}^{m_2} \sum_{\alpha_1,\cdots,\alpha_n=1}^{m_1} \left(\prod_{i=1}^n w'_{j_i}\right)\left(\prod_{i=1}^n w_{\alpha_i}\right) \\&~~~~g_n(r_{j_1},\cdots,r_{j_n},s_{\alpha_1},\cdots,s_{\alpha_n}).
\end{aligned}
\end{equation}
\else
\begin{equation}
	\begin{aligned}
 V_m^n(g_n) & = \sum_{j_1,\cdots,j_n=1}^{m_2} \sum_{\alpha_1,\cdots,\alpha_n=1}^{m_1} \left(\prod_{i=1}^n w_{j_i}\right)\left(\prod_{i=1}^n w_{\alpha_i}\right)g_n(r_{j_1},\cdots,r_{j_n},s_{\alpha_1},\cdots,s_{\alpha_n})
\end{aligned}
\end{equation}
\fi

To make the derivation easier to understand, we will first discuss the scenario when $m_1=m_2 = m$, where the transceivers are under same discretization order, $w'_i = w_i$ and $s_i=r_i$. Under this scenario we have
\begin{equation}
	\begin{aligned}
V_m^n(g_n)= \sum_{j_1,\cdots,j_{2n}=1}^m \left(\prod_{i=1}^{2n} w_{j_i}\right) g_n(r_{j_1},\cdots,r_{j_{2n}}).
\end{aligned}
\end{equation}

	The difference between $d(z)$ and $d_V(z)$ is 
	\ifx\onecol\undefined
	\begin{equation}
		\begin{aligned}
	&~~~~d(z)-d_V(z) \\&= \sum_{n=1}^{\infty}\frac{z^n}{n!}\Bigg( V^n_m(g_n)-\int_{[0,l]^n}K_n(x_1,\cdots,x_n){\rm d}x_1\cdots{\rm d}x_n \Bigg)
	\\& = \sum_{n=1}^{\infty}\frac{z^n}{n!}\Bigg( V^n_m(g_n)-\int_{[0,l]^{2n}} g_n(x_1,\cdots,x_{2n}){\rm d}x_1\cdots{\rm d}x_{2n} \Bigg)
\end{aligned}
\end{equation}
\else
\begin{equation}
	\begin{aligned}
d(z)-d_V(z) &= \sum_{n=1}^{\infty}\frac{z^n}{n!}\left( V^n_m(g_n)-\int_{[0,l]^n}K_n(x_1,\cdots,x_n){\rm d}x_1\cdots{\rm d}x_n \right)
\\& = \sum_{n=1}^{\infty}\frac{z^n}{n!}\left( V^n_m(g_n)-\int_{[0,l]^{2n}} g_n(x_1,\cdots,x_{2n}){\rm d}x_1\cdots{\rm d}x_{2n} \right).
\end{aligned}
\end{equation}
\fi

Note that $V^n_m(g_n)$ is the numerical discretization of the function $g_n$ with $2n$ variables,
we can bound $V^n_m(g_n)-\int_{[a,b]^{2n}} g_n(x_1,\cdots,x_{2n}){\rm d}x_1\cdots{\rm d}x_{2n}$ by using the multivariate numerical integration error bound:
\ifx\onecol\undefined
\begin{equation}
	\begin{aligned}
		&\left|  V^n_m(g_n)-\int_{[0,l]^{2n}} g_n(x_1,\cdots,x_{2n}){\rm d}x_1\cdots{\rm d}x_{2n}  \right| \\&\leqslant l^{2n-1} \sum_{i=1}^{2n}E_i,
	\end{aligned}
\end{equation}
\else
\begin{equation}
	\begin{aligned}
		\left|  V^n_m(g_n)-\int_{[0,l]^{2n}} g_n(x_1,\cdots,x_{2n}){\rm d}x_1\cdots{\rm d}x_{2n}  \right| \leqslant l^{2n-1} \sum_{i=1}^{2n}E_i,
	\end{aligned}
\end{equation}
\fi
where 
\begin{equation}
	\begin{aligned}
		E_i = \left| Q_i(g_n;x_i) - \int_0^l g_n(x_1,\cdots,x_{2n}){\rm d}x_i \right|.
	\end{aligned}
\end{equation}
According to the $m$-point composite midpoint quadrature rule, we have
\ifx\onecol\undefined
\begin{equation}
	\begin{aligned}
		&~~\left| Q_i(g_n;x_i) - \int_0^l g_n(x_1,\cdots,x_{2n}){\rm d}x_i \right| \\&\leqslant \frac{l^{3}}{24m^2} \left\|    \frac{\partial^2 g_n}{\partial x_i^2}\right\|_{L^{\infty}((0,l)^{2n})},
	\end{aligned}
	\label{equ_Q_m1_equal_m2}
\end{equation}
\else
\begin{equation}
	\begin{aligned}
		&~~\left| Q_i(g_n;x_i) - \int_0^l g_n(x_1,\cdots,x_{2n}){\rm d}x_i \right| \leqslant \frac{l^{3}}{24m^2} \left\|    \frac{\partial^2 g_n}{\partial x_i^2}\right\|_{L^{\infty}((0,l)^{2n})},
	\end{aligned}
	\label{equ_Q_m1_equal_m2}
\end{equation}
\fi

From the Hadamard's inequality \cite{meyer2000matrix}, we can further bound it by
\ifx\onecol\undefined
\begin{equation}
	\begin{aligned}
		&~~\left| V_m^n(g_n) - \int_{[a,b]^{2n}} g_n(x_1,\cdots,x_{2n}){\rm d}x_1\cdots {\rm d}x_{2n}\right| \\&\leqslant 2n \frac{l^{2n+2}}{24m^2} \underset{j}{\rm max}\left\|    \frac{\partial^2 g_n}{\partial x_j^2}\right\|_{L^{\infty}((0,l)^{2n})}
		\\& \leqslant  2n\frac{l^{2n+2}}{24m^2} {\rm max}\bigggl( n^{n/2} \left( \left\|    \frac{\partial^2 g(x,y,z)}{\partial z^2}\right\|_{L^{\infty}((0,l)^{3})}\right)^n, \\&~~~4 n^{n/2} \left( \underset{i+j\leqslant 2}{{\rm max}}\left\| \frac{\partial_{x}^i \partial_{y}^j g(x,y,z)}{\partial x^i\partial y^j} \right\|_{L^{\infty}((0,l)^3)} \right)^n\bigggl)
		\\&\leqslant \frac{l^{2n+2}}{3m^2}n^{(n+2)/2} \left( \underset{i+j+k\leqslant 2}{\rm max}\left\| \frac{\partial_{x}^i \partial_{y}^j \partial_{z}^k g(x,y,z)}{\partial x^i\partial y^j\partial z^k} \right\|_{L^{\infty}((0,l)^3)} \right)^n.
	\end{aligned}
	\label{equ_m1_equal_m2}
\end{equation}
\else
\begin{equation}
	\begin{aligned}
		&~~~\left| V_m^n(g_n) - \int_{[a,b]^{2n}} g_n(x_1,\cdots,x_{2n}){\rm d}x_1\cdots {\rm d}x_{2n}\right| \leqslant 2n \frac{l^{2n+2}}{24m^2} \underset{j}{\rm max}\left\|    \frac{\partial^2 g_n}{\partial x_j^2}\right\|_{L^{\infty}((0,l)^{2n})}
		\\& \leqslant  2n\frac{l^{2n+2}}{24m^2} {\rm max}\bigggl( n^{n/2} \left( \left\|    \frac{\partial^2 g(x,y,z)}{\partial z^2}\right\|_{L^{\infty}((0,l)^{3})}\right)^n, 4 n^{n/2} \left( \underset{i+j\leqslant 2}{{\rm max}}\left\| \frac{\partial_{x}^i \partial_{y}^j g(x,y,z)}{\partial x^i\partial y^j} \right\|_{L^{\infty}((0,l)^3)} \right)^n\bigggl)
		\\&\leqslant \frac{l^{2n+2}}{3m^2}n^{(n+2)/2} \left( \underset{i+j+k\leqslant 2}{\rm max}\left\| \frac{\partial_{x}^i \partial_{y}^j \partial_{z}^k g(x,y,z)}{\partial x^i\partial y^j\partial z^k} \right\|_{L^{\infty}((0,l)^3)} \right)^n.
	\end{aligned}
	\label{equ_m1_equal_m2}
\end{equation}
\fi 

Similar to {\bf Lemma 3}, we know that when $m_1=m_2 = m$ and $m \rightarrow \infty$, $|I_2-I_0^{'}|$ converges to 0, and the error is at most inverse proportional to $m^2$. 

For the scenario when $m_1 \neq m_2$, we can extend the result of (\ref{equ_m1_equal_m2}) similar to {\bf Lemma 4}. Notice that 
\ifx\onecol\undefined
\begin{equation}
	\begin{aligned}
		&~~~~\left| Q_i(g_n;x_i) - \int_0^l g_n(x_1,\cdots,x_{2n}){\rm d}x_i \right| \leqslant
		\\&\left\{ 
\begin{array} {lcr}
		  \frac{l^{3}}{24{m_2}^2} \left\|    \frac{\partial^2 g_n}{\partial x_i^2}\right\|_{L^{\infty}((0,l)^{2n})},  & 1\leqslant i \leqslant n 
		 \\ \frac{l^{3}}{24{m_1}^2} \left\|    \frac{\partial^2 g_n}{\partial x_i^2}\right\|_{L^{\infty}((0,l)^{2n})},  & {n+1}\leqslant i \leqslant 2n
		\end{array} 
		\right.
	\end{aligned}
\end{equation}
\else
\begin{equation}
	\begin{aligned}
		\left| Q_i(g_n;x_i) - \int_0^l g_n(x_1,\cdots,x_{2n}){\rm d}x_i \right| \leqslant
		\left\{ 
\begin{array} {lcr}
		  \frac{l^{3}}{24{m_2}^2} \left\|    \frac{\partial^2 g_n}{\partial x_i^2}\right\|_{L^{\infty}((0,l)^{2n})},  & 1\leqslant i \leqslant n 
		 \\ \frac{l^{3}}{24{m_1}^2} \left\|    \frac{\partial^2 g_n}{\partial x_i^2}\right\|_{L^{\infty}((0,l)^{2n})},  & {n+1}\leqslant i \leqslant 2n
		\end{array} 
		\right.
	\end{aligned}
\end{equation}
\fi
Comparing with (\ref{equ_Q_m1_equal_m2}) and (\ref{equ_m1_equal_m2}), it is trivial to obtain that  
\ifx\onecol\undefined
\begin{equation}
	\begin{aligned}
		&~~~\left| V_m^n(g_n) - \int_{[a,b]^{2n}} g_n(x_1,\cdots,x_{2n}){\rm d}x_1\cdots {\rm d}x_{2n}\right| 
		\\&\leqslant \frac{l^{2n+2}}{3({\rm min}(m_1,m_2))^2}n^{(n+2)/2} \\&\left( \underset{i+j+k\leqslant 2}{\rm max}\left\| \frac{\partial_{x}^i \partial_{y}^j \partial_{z}^k g(x,y,z)}{\partial x^i\partial y^j\partial z^k} \right\|_{L^{\infty}((0,l)^3)} \right)^n.
		\label{equ_m1_unequal_m2}
	\end{aligned}
\end{equation}
\else
\begin{equation}
	\begin{aligned}
		&~~~\left| V_m^n(g_n) - \int_{[a,b]^{2n}} g_n(x_1,\cdots,x_{2n}){\rm d}x_1\cdots {\rm d}x_{2n}\right| 
		\\&\leqslant \frac{l^{2n+2}}{3({\rm min}(m_1,m_2))^2}n^{(n+2)/2} \left( \underset{i+j+k\leqslant 2}{\rm max}\left\| \frac{\partial_{x}^i \partial_{y}^j \partial_{z}^k g(x,y,z)}{\partial x^i\partial y^j\partial z^k} \right\|_{L^{\infty}((0,l)^3)} \right)^n.
		\label{equ_m1_unequal_m2}
	\end{aligned}
\end{equation}
\fi
Then, we know that $|I_2-I_0^{'}|$ converges to 0, and the error is at most inverse proportional to $({\rm min}(m_1,m_2))^2$.
\end{IEEEproof}
Therefore, we have {\bf Theorem 2}:
\begin{theorem}
	The mutual information $I_2$ that can be obtained from the discrete transceivers converges to the mutual information $I_0$ that can be obtained from the continuous transceivers when the number of antennas in the discrete transceivers increases. The difference $|I_0-I_2|$ is at most inverse-proportional to the square of $m$, where $m = {\rm min}(m_1,m_2)$.
\end{theorem}

\begin{remark}
	Insight about the sampling numbers: {\bf Theorem 2} reveals the convergence of $|I_0-I_2|$ with respect to $m = {\rm min}(m_1,m_2)$. It shows that the mutual information of the discrete MIMO system depends on both sampling numbers $m_1$ and $m_2$ of the transceivers. They have some symmetry in (\ref{equ_n0_m1m2_n2}) and (\ref{equ_m1_unequal_m2}), which we will show further in the numerical analysis. Moreover, if $m_2>m_1$, then increasing $m_2$ while $m_1$ is fixed will not apparently increase the mutual information that can be obtained. Therefore, for mutual information, there is a short board effect in the discretization of transceiver with dense antennas.
\end{remark}

\begin{remark}
	Extension to other scenarios with power allocation: similar to {\bf Remark 3} in Section. \ref{sec_discrete_receiver}, the convergence analysis in this section is not limited to the scenario with equal power allocation. For arbitrary analytic function $R_J(s,s')$, the convergence of $|I_0-I_2|$ can be obtained. Instead of discretizing $\int G(r,z)G^{*}(r',z){\rm d}z$ to $\sum_i G(r,r_i)G^{*}(r',r_i)$, we will discretize $\iint G(r,z)R_J(z,z')G^{*}(r',z'){\rm d}z{\rm d}z'$ to $\sum_{i,j} G(r,r_i)R_J(r_i,r_j)G^{*}(r',r_j)$ in the extended scenarios with power allocation schemes. Then, instead of $g(x,y,z)$ we need a four-variable function $h(x,y,z,\omega):= G(x,z)R_J(z,\omega)G^{*}(y,\omega)$ and the derivation procedure of the convergence has no essential difference with {\bf Theorem 2} if the smoothness of $h(x,y,z,\omega)$ is guaranteed. 
\end{remark}

\subsection{Numerical analysis about the mutual information}
\ifx\onecol\undefined
\begin{figure}
	\centering 
	\includegraphics[width=0.5\textwidth]{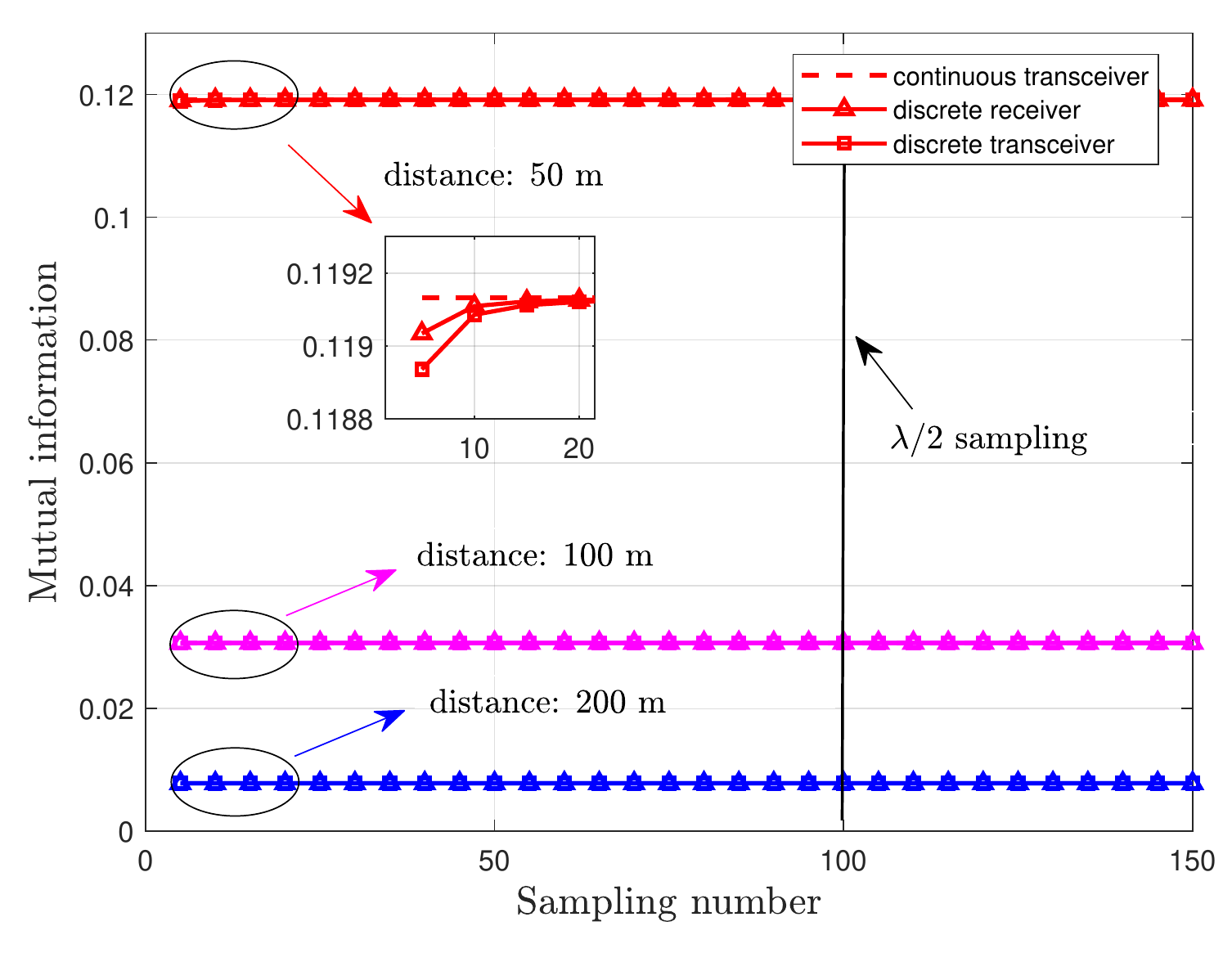} 
	\caption{The mutual information variation with different sampling numbers. The mutual information that corresponds to the three models with continuous and discrete transceivers is plotted. The distance between the transceivers is large.} 
	\label{fig_discrete_transceiver2}
\end{figure}
\else
\begin{figure*}
	\centering 
	\includegraphics[width=0.7\textwidth]{figs1/discrete_transceiver2.pdf} 
	\caption{The mutual information variation with different sampling numbers. The mutual information that corresponds to the three models with continuous and discrete transceivers is plotted. The distance between the transceivers is large.} 
	\label{fig_discrete_transceiver2}
\end{figure*}
\fi
In this subsection, we will verify the correctness of the convergence analysis in the above subsection by simulations. The length $l$ of the transceivers is fixed to 2\,$\rm m$. We have plotted the mutual information of the three models: continuous transceiver, continuous transmitter and discrete receiver, and discrete transceiver. The wavelength of the electromagnetic field is fixed to 0.04\,$\rm m$, which corresponds to the frequency of 7.5\,$\rm GHz$.
First we will show the scenarios when the distance between the transceivers is large. The distance between the transceivers varies from 50\,$\rm m$ to 200\,$\rm m$. 

First we will discuss the scenario where the transceivers are both discretized to $m$ point antennas.
The simulation results are shown in Fig. \ref{fig_discrete_transceiver2}.
From the simulation results we find that the mutual information nearly keeps the same when the sampling number increases. The reason for this phenomenon is that the DoF of the channel is nearly inverse-proportional to the distance between transceivers. For example, the DoF when the distance equals 50\,$\rm m$ can be approximated by $l^2/(d\lambda) = 2$, which means that when the sampling number is 5, the multiplexing gain is almost fully explored. Therefore, for large distances between transceivers, the dominant limitation is the channel DoF, which means that the suboptimal performance can be achieved by sampling sparser than half-wavelength. 

\ifx\onecol\undefined
\begin{figure}
	\centering 
	\includegraphics[width=0.5\textwidth]{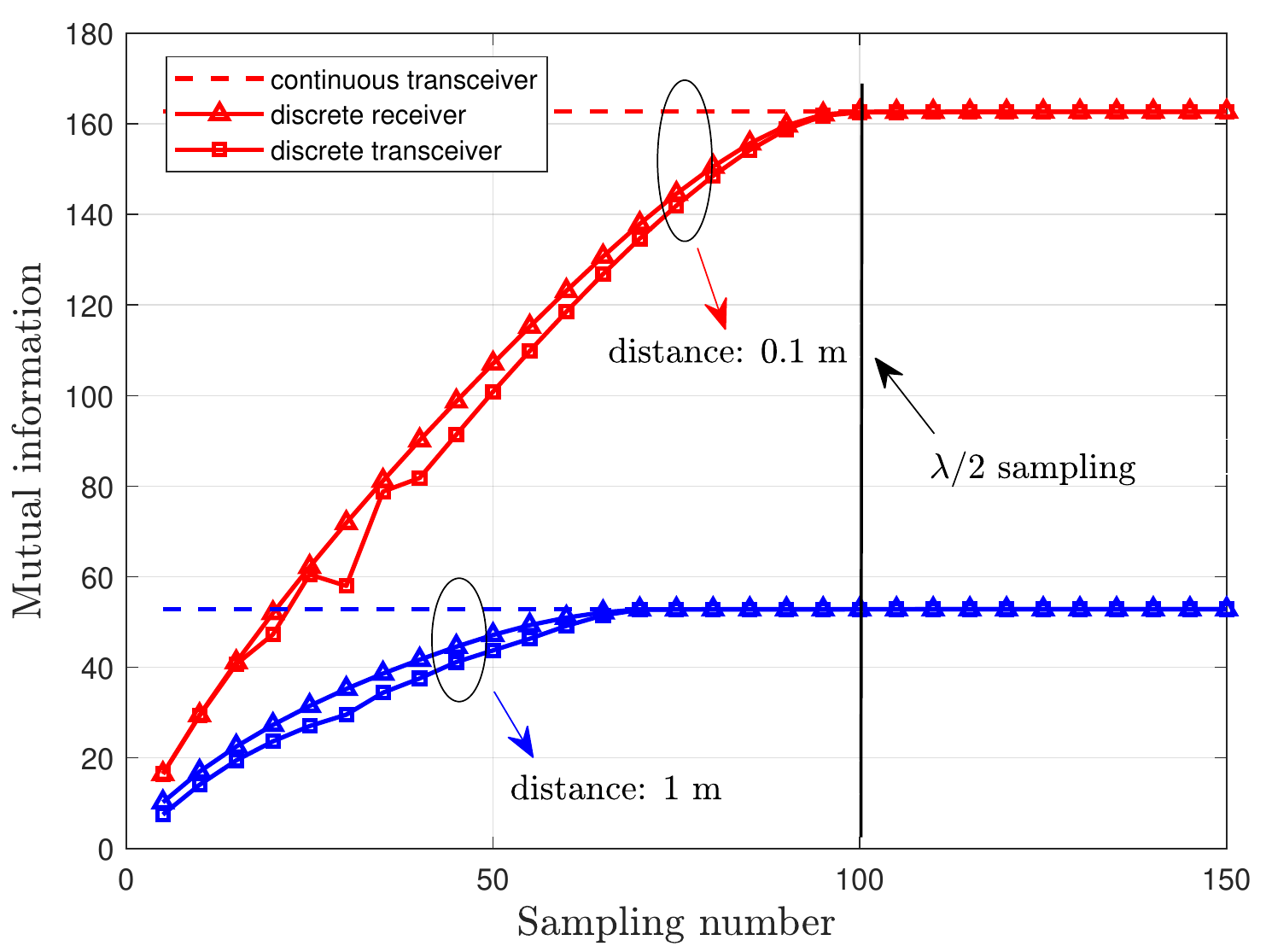} 
	\caption{The mutual information as a function of the sampling number. The mutual information values that correspond to the three models with continuous and discrete transceivers are plotted. The distance between the transceivers is small.} 
	\label{fig_discrete_transceiver1}
\end{figure}
\else
\begin{figure*}
	\centering 
	\includegraphics[width=0.7\linewidth]{figs1/discrete_transceiver1.pdf} 
	\caption{The mutual information as a function of the sampling number. The mutual information values that correspond to the three models with continuous and discrete transceivers are plotted. The distance between the transceivers is small.} 
	\label{fig_discrete_transceiver1}
\end{figure*}
\fi

Moreover, we have shown the variation of the mutual information with the sampling number when the distance between transceivers is small. In Fig. \ref{fig_discrete_transceiver1} the distance between the transceiver is 0.1\,$\rm m$ and 1\,$\rm m$. We can find that when the distance decreases, the dense sampling of the transceivers becomes important to fully explore the limit of the mutual information. However, the half-wavelength sampling of the transceivers still achieves suboptimal performance, which means that denser sampling schemes are not necessary.

\ifx\onecol\undefined
\begin{figure}
	\centering 
	\includegraphics[width=0.5\textwidth]{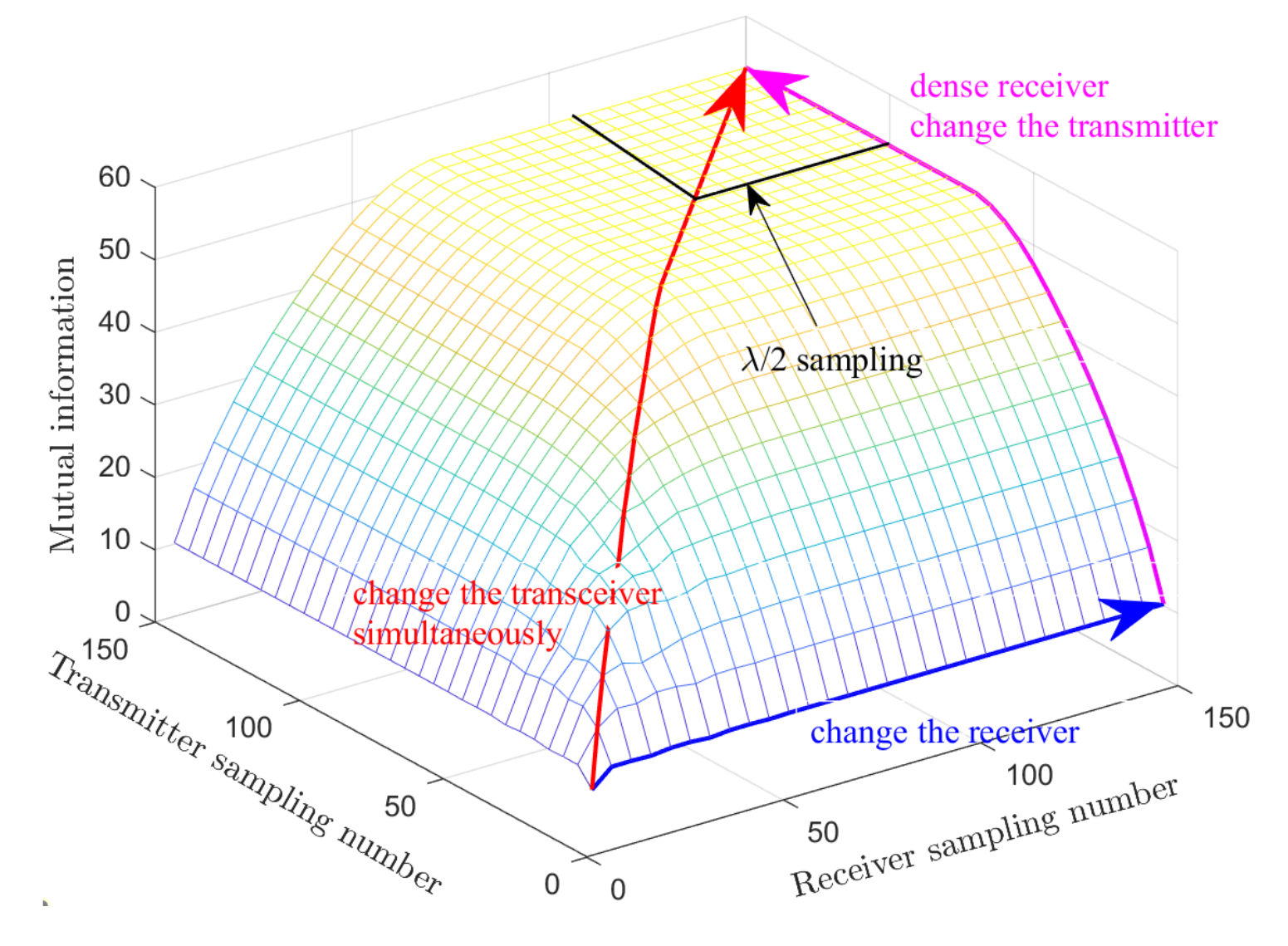} 
	\caption{The mutual information as a function of the sampling numbers of the transmitter and the receiver.} 
	\label{fig_unequal_sampling}
\end{figure}
\else
\begin{figure*}
	\centering 
	\includegraphics[width=0.7\linewidth]{figs1/unequal_sampling.pdf} 
	\caption{The mutual information as a function of the sampling numbers of the transmitter and the receiver.} 
	\label{fig_unequal_sampling}
\end{figure*}
\fi

Finally, in Fig. \ref{fig_unequal_sampling}, we show the change of mutual information with respect to both $m_1$ and $m_2$, which are the sampling number of the transceivers. The distance $d$ is 1\,$\rm m$. We can find that, as we have predicted in {\bf Remark 5}, the figure is approximately symmetric for $m_1$ and $m_2$. When we change the sampling numbers of the transceivers simultaneously, the mutual information increases obviously as the red line shows. When the $m_2$ is large enough in the rosy line, which corresponds to dense antennas in the receiver, the mutual information will also increase with $m_1$. When $m_1$ or $m_2$ is small, which corresponds to sparse antennas, increasing the other sampling number only has a slight improvement on the mutual information, as the blue line shows. Therefore, the shortboard effect determines how the sampling numbers influence the mutual information. Moreover, the near-optimality of the half-wavelength sampling transceivers is also shown in Fig. \ref{fig_unequal_sampling}.

\section{Conclusion}
\label{sec_conclusion}
In this paper, we proposed a comparison scheme between continuous and discrete MIMO systems which is based on a precise non-asymptotic analysis framework. Three information-theoretic models of the continuous and discrete transceivers were built, with the first model corresponds to the fully continuous electromagnetic information theory model, and the third model corresponds to the matrix-vector MIMO model. We proposed physically consistent SNR control schemes to ensure the fairness of the comparison, and proved that the mutual information between discrete MIMO transceivers converges to that between continuous electromagnetic transceivers. Numerical results verified the theoretical analysis and showed the near-optimality of the half-wavelength sampling scheme. 

Further works can be done by extending the linear transceivers to rectangular or other two-dimensional transceivers for generality. The analysis based on the capacity after water-filling of the mutual information also remains to be explored.

\section*{Appendix A \\ Extension of Lemma 2}
In this part, we will show how {\bf Lemma 2} can be extended to the integral operators with several variables, which is key to the convergence proof for other scenarios. Here we use the two-dimensional transceivers as an example, where the transmitter is in the region $s \in [a,b],s' \in [c,d]$ and the receiver is in the region $r\in [a,b],r' \in [c,d]$. The integral operator $T$ is $T:= \phi(x_1,x_2)\rightarrow \int_a^b\int_c^d K(x_1,x_2,y_1,y_2)\phi(y_1,y_2){\rm d}y_1{\rm d}y_2$. The kernel $K$ comes from the autocorrelation function of the electromagnetic field, where $K(x_1,x_2,y_1,y_2) = \mathbb{E}[E(x_1,x_2)E^{*}(y_1,y_2)]$. We denote ${\bf K}$ by the four-dimensional tensor ${\bf K}_{i_1,j_1,i_2,j_2} = w_{i_2}w'_{j_2} K(r_{i_1},r'_{j_1},r_{i_2},r'_{j_2})$. Here $w_{i_2}$ and $w'_{j_2}$ denote the antenna spacing in the two directions of the receiver. We view this tensor as a discrete operator that projects a matrix $C(r_{i_2},r'_{j_2})_{i_2,j_2=1}^{m,m'}$ into $D(r_{i_1},r'_{j_1})_{i_1,j_1=1}^{m,m'}$.
From the definition of determinant using exterior algebra \cite{simon2005trace}, we have 
\begin{equation}
	{\rm det}(1+zA) = \sum_{n=0}^{\infty} z^n{\rm tr}\left( \Lambda^n (A) \right),
	\label{equ_exterior_algebra}
\end{equation}  
where $A:\mathcal{H}\rightarrow \mathcal{H} $ is a trace-class operator, $\mathcal{H}$ is a Hilbert space, $\Lambda^n (A) = \underset{{\rm n times}}{\underbrace{A \otimes \cdots \otimes A}}$. The Hilbert space $\mathcal{H}$ is spanned by the orthonormal basis $\{\phi_i \}$.
Denote the wedge product $\phi_{i_1} \wedge \cdots \wedge \phi_{i_n} = \frac{1}{\sqrt{n!}}\sum_{\pi \in \sigma_n} (-1)^{\pi} \phi_{i_{\pi(1)}} \otimes \cdots \otimes \phi_{i_{\pi(n)}}$. $\Lambda^n (\mathcal{H})$ is a Hilbert space spanned by $\phi_{i_1} \wedge \cdots \wedge \phi_{i_n}$. From the definition of $\phi_{i_1} \wedge \cdots \wedge \phi_{i_n}$ we know that exchange $\phi_{i_\alpha}$ and $\phi_{i_\beta}$ does not form a new base function. 

For the tensor ${\bf K}$, we denote the bases of its domain of definition $\mathcal{H}$ by ${\bf C}_{\bf i} = {\bf C}_{i,j} = C({\mathbbm 1}_{i_1=i},{\mathbbm 1}_{j_1=j})_{i_1,j_1=1}^{m,m'}$. The order of the composite index ${\bf i} =(i,j)$ is determined by ${\bf i}_1>{\bf i}_2$ when $i_1>i_2$ or $i_1=i_2,j_1>j_2$. According to the definition of trace, we have 
\ifx\onecol\undefined
\begin{equation}
	\begin{aligned}
	{\rm tr}\left( \Lambda^n ({\bf K}) \right)= &\sum_{{\bf i}_1<\cdots<{\bf i}_n} \langle {\bf C}_{{\bf i}_1} \wedge \cdots \wedge {\bf C}_{{\bf i}_n}| \\&\Lambda^n ({\bf K}) \left( {\bf C}_{{\bf i}_1} \wedge \cdots \wedge {\bf C}_{{\bf i}_n} \right) \rangle
	\\= &\frac{1}{n!} \sum_{{\bf i}_1,\cdots,{\bf i}_n=1}^{(m,m')} \langle {\bf C}_{{\bf i}_1} \wedge \cdots \wedge {\bf C}_{{\bf i}_n}| \\& \Lambda^n ({\bf K}) \left( {\bf C}_{{\bf i}_1} \wedge \cdots \wedge {\bf C}_{{\bf i}_n} \right) \rangle.
	\end{aligned}
\end{equation}
\else
\begin{equation}
	\begin{aligned}
	{\rm tr}\left( \Lambda^n ({\bf K}) \right)& = \sum_{{\bf i}_1<\cdots<{\bf i}_n} \langle {\bf C}_{{\bf i}_1} \wedge \cdots \wedge {\bf C}_{{\bf i}_n}| \Lambda^n ({\bf K}) \left( {\bf C}_{{\bf i}_1} \wedge \cdots \wedge {\bf C}_{{\bf i}_n} \right) \rangle
	\\& = \frac{1}{n!} \sum_{{\bf i}_1,\cdots,{\bf i}_n=1}^{(m,m')} \langle {\bf C}_{{\bf i}_1} \wedge \cdots \wedge {\bf C}_{{\bf i}_n}| \Lambda^n ({\bf K}) \left( {\bf C}_{{\bf i}_1} \wedge \cdots \wedge {\bf C}_{{\bf i}_n} \right) \rangle.
	\end{aligned}
\end{equation} 
\fi

By using the definition of the wedge product $\wedge$, we have
\ifx\onecol\undefined
\begin{equation}
	\begin{aligned}
	&~~~~{\rm tr}\left( \Lambda^n ({\bf K}) \right)\\& = \frac{1}{n!} \frac{1}{n!} \sum_{{\bf i}_1,\cdots,{\bf i}_n=1}^{(m,m')} \sum_{\pi_1 \in \sigma_n} \sum_{\pi_2 \in \sigma_n} (-1)^{\pi_1+\pi_2} \\&~~~~ \prod_k \left( \langle {\bf C}_{{\bf i}_{\pi_1(k)}} | {\bf K}{\bf C}_{{\bf i}_{\pi_2(k)}} \rangle \right)
	\\& = \frac{1}{n!}\sum_{{\bf i}_1,\cdots,{\bf i}_n=1}^{(m,m')} \sum_{\pi \in \sigma_n} (-1)^{\pi} \prod_k \left( \langle {\bf C}_{{\bf i}_k} | {\bf K}{\bf C}_{{\bf i}_{\pi(k)}} \rangle \right)
	\\& = \frac{1}{n!}\sum_{{i}_1,\cdots,{i}_n=1}^{m} \sum_{{j}_1,\cdots,{j}_n=1}^{m'} \sum_{\pi \in \sigma_n} (-1)^{\pi} \prod_k {\bf K}_{i_k,j_k,i_{\pi(k)},j_{\pi(k)}}.
	\end{aligned}
	\label{equ_tr_K_2variable}
\end{equation}
\else
\begin{equation}
	\begin{aligned}
	{\rm tr}\left( \Lambda^n ({\bf K}) \right)& = \frac{1}{n!} \frac{1}{n!} \sum_{{\bf i}_1,\cdots,{\bf i}_n=1}^{(m,m')} \sum_{\pi_1 \in \sigma_n} \sum_{\pi_2 \in \sigma_n} (-1)^{\pi_1+\pi_2} \prod_k \left( \langle {\bf C}_{{\bf i}_{\pi_1(k)}} | {\bf K}{\bf C}_{{\bf i}_{\pi_2(k)}} \rangle \right)
	\\& = \frac{1}{n!}\sum_{{\bf i}_1,\cdots,{\bf i}_n=1}^{(m,m')} \sum_{\pi \in \sigma_n} (-1)^{\pi} \prod_k \left( \langle {\bf C}_{{\bf i}_k} | {\bf K}{\bf C}_{{\bf i}_{\pi(k)}} \rangle \right)
	\\& = \frac{1}{n!}\sum_{{i}_1,\cdots,{i}_n=1}^{m} \sum_{{j}_1,\cdots,{j}_n=1}^{m'} \sum_{\pi \in \sigma_n} (-1)^{\pi} \prod_k {\bf K}_{i_k,j_k,i_{\pi(k)},j_{\pi(k)}}.
	\end{aligned}
	\label{equ_tr_K_2variable}
\end{equation}
\fi

For the integral operator $T$ with two variables, inspired by the derivation procedure of integral operator with one variable in \cite{simon2005trace}, first we note that $\langle \phi_{i_1} \wedge \cdots \wedge \phi_{i_n}| \Lambda^n (T) \left( \phi_{i_1} \wedge \cdots \wedge \phi_{i_n} \right) \rangle = \langle Q_n\left(\phi_{i_1} \otimes \cdots \otimes \phi_{i_n} \right)| \Lambda^n (T)Q_n\left( \phi_{i_1} \otimes \cdots \otimes \phi_{i_n} \right) \rangle$, where $Q_n$ is the projection from $\otimes^nL^2((a,b)\times (c,d))$ to $\Lambda^nL^2((a,b)\times(c,d))$. It is obvious that $\langle Q_n \left( \phi_{i_1} \otimes \cdots \otimes \phi_{i_n} \right) | \psi_{i_1} \otimes \cdots \otimes \psi_{i_n} \rangle = \langle \phi_{i_1} \otimes \cdots \otimes \phi_{i_n} | Q_n \left( \psi_{i_1} \otimes \cdots \otimes \psi_{i_n} \right) \rangle$, since
\ifx\onecol\undefined
\begin{equation}
	\begin{aligned}
	&~~~~\langle Q_n \left( \phi_{i_1} \otimes \cdots \otimes \phi_{i_n} \right) | \psi_{i_1} \otimes \cdots \otimes \psi_{i_n} \rangle \\&= \frac{1}{\sqrt{n!}} \sum_{\pi \in \sigma_n}(-1)^\pi \langle \phi_{i_{\pi(1)}} \otimes \cdots \otimes \phi_{i_{\pi(n)}} | \psi_{i_1} \otimes \cdots \otimes \psi_{i_n} \rangle
	\\& = \frac{1}{\sqrt{n!}} \sum_{\pi^{-1} \in \sigma_n}(-1)^\pi \langle \phi_{i_{1}} \otimes \cdots \otimes \phi_{i_{n}} | \\&~~~~ \psi_{i_{\pi^{-1}(1)}} \otimes \cdots \otimes \psi_{i_{\pi^{-1}(n)}} \rangle
	\\& = \langle \phi_{i_1} \otimes \cdots \otimes \phi_{i_n} | Q_n \left( \psi_{i_1} \otimes \cdots \otimes \psi_{i_n} \right) \rangle.
	\end{aligned}
\end{equation}
\else
\begin{equation}
	\begin{aligned}
	\langle Q_n \left( \phi_{i_1} \otimes \cdots \otimes \phi_{i_n} \right) | \psi_{i_1} \otimes \cdots \otimes \psi_{i_n} \rangle &= \frac{1}{\sqrt{n!}} \sum_{\pi \in \sigma_n}(-1)^\pi \langle \phi_{i_{\pi(1)}} \otimes \cdots \otimes \phi_{i_{\pi(n)}} | \psi_{i_1} \otimes \cdots \otimes \psi_{i_n} \rangle
	\\& = \frac{1}{\sqrt{n!}} \sum_{\pi^{-1} \in \sigma_n}(-1)^\pi \langle \phi_{i_{1}} \otimes \cdots \otimes \phi_{i_{n}} | \psi_{i_{\pi^{-1}(1)}} \otimes \cdots \otimes \psi_{i_{\pi^{-1}(n)}} \rangle
	\\& = \langle \phi_{i_1} \otimes \cdots \otimes \phi_{i_n} | Q_n \left( \psi_{i_1} \otimes \cdots \otimes \psi_{i_n} \right) \rangle.
	\end{aligned}
\end{equation}
\fi

Therefore, the trace of $\Lambda^n (T)$ equals the trace of $Q_n\Lambda^n (T) Q_n $, which differ by only the factor $1/n!$. The reason is that every base function $\underset{i_1 < \cdots < i_n}{\underbrace{\phi_{i_1} \wedge \cdots \wedge \phi_{i_n}}}$ in $\Lambda^n \mathcal{H}$ corresponds to $n!$ base functions $\underset{\pi \in \sigma_n}{\underbrace{\phi_{i_{\pi(1)}} \otimes \cdots \otimes \phi_{i_{\pi(n)}}}}$ in $\otimes^n \mathcal{H}$.

For $Q_n \Lambda^n(T) Q_n$, we have 
\ifx\onecol\undefined
\begin{equation}
	\begin{aligned}
		&~~~~Q_n \Lambda^n(T) Q_n\left( \phi_1 \otimes \cdots \otimes \phi_n \right) \\&= \frac{1}{\sqrt{n!}} Q_n \Lambda^n(T) \left( \sum_{\pi_1 \in \sigma_n} (-1)^{\pi_1} \phi_{\pi_1(1)} \otimes \cdots \otimes \phi_{\pi_1(n)}\right)
		\\& = \frac{1}{\sqrt{n!}} Q_n \left( \sum_{\pi_1 \in \sigma_n} (-1)^{\pi_1} T\phi_{\pi_1(1)} \otimes \cdots \otimes T\phi_{\pi_1(n)}\right)
		\\& = \frac{1}{\sqrt{n!}}\sum_{\pi_1 \in \sigma_n} (-1)^{\pi_1} T\phi_{\pi_1(1)} \wedge \cdots \wedge T\phi_{\pi_1(n)}
		\\& = \frac{1}{\sqrt{n!}}\frac{1}{\sqrt{n!}} \sum_{\pi_1 \in \sigma_n}\sum_{\pi_2 \in \sigma_n} (-1)^{\pi_1+\pi_2} \\&~~~~T\phi_{\pi_2\pi_1(1)}\otimes \cdots \otimes T\phi_{\pi_2\pi_1(n)}
		\\& = \sum_{\pi \in \sigma_n}(-1)^\pi T\phi_{\pi(1)}\otimes \cdots \otimes T\phi_{\pi(n)}
		\\& = \sum_{\pi \in \sigma_n}(-1)^\pi \\&~~~~\int_a^b\int_c^d K(x_{11},x_{21},y_{11},y_{21}) \phi_{\pi(1)}(y_{11},y_{21}) {\rm d}y_{11} {\rm d}y_{21} \cdots \\&~~~~ \int_a^b \int_c^d K(x_{1n},x_{2n},y_{1n},y_{2n}) \phi_{\pi(n)}(y_{1n},y_{2n}){\rm d}y_{1n}{\rm d}y_{2n}
		\\& = \int_{((a,b)\times (c,d))^{n}}\sum_{\pi \in \sigma_n}(-1)^\pi K(x_{11},x_{21},y_{1\pi(1)},y_{2\pi(1)})\cdots \\&~~~~K(x_{1n},x_{2n},y_{1\pi(n)},y_{2\pi(n)})  \phi_1(y_{11},y_{21}) \cdots \phi_n(y_{1n},y_{2n}) \\&~~~~{\rm d}y_{11} \cdots {\rm d}y_{2n}. 
	\end{aligned}
\end{equation}
\else
\begin{equation}
	\begin{aligned}
		&~~~~Q_n \Lambda^n(T) Q_n\left( \phi_1 \otimes \cdots \otimes \phi_n \right) = \frac{1}{\sqrt{n!}} Q_n \Lambda^n(T) \left( \sum_{\pi_1 \in \sigma_n} (-1)^{\pi_1} \phi_{\pi_1(1)} \otimes \cdots \otimes \phi_{\pi_1(n)}\right)
		\\& = \frac{1}{\sqrt{n!}} Q_n \left( \sum_{\pi_1 \in \sigma_n} (-1)^{\pi_1} T\phi_{\pi_1(1)} \otimes \cdots \otimes T\phi_{\pi_1(n)}\right)
		\\& = \frac{1}{\sqrt{n!}}\sum_{\pi_1 \in \sigma_n} (-1)^{\pi_1} T\phi_{\pi_1(1)} \wedge \cdots \wedge T\phi_{\pi_1(n)}
		\\& = \frac{1}{\sqrt{n!}}\frac{1}{\sqrt{n!}} \sum_{\pi_1 \in \sigma_n}\sum_{\pi_2 \in \sigma_n} (-1)^{\pi_1+\pi_2} T\phi_{\pi_2\pi_1(1)}\otimes \cdots \otimes T\phi_{\pi_2\pi_1(n)}
		\\& = \sum_{\pi \in \sigma_n}(-1)^\pi T\phi_{\pi(1)}\otimes \cdots \otimes T\phi_{\pi(n)}
		\\& = \sum_{\pi \in \sigma_n}(-1)^\pi \int_a^b\int_c^d K(x_{11},x_{21},y_{11},y_{21}) \phi_{\pi(1)}(y_{11},y_{21}) {\rm d}y_{11} {\rm d}y_{21} \cdots \\&~~~~ \int_a^b \int_c^d K(x_{1n},x_{2n},y_{1n},y_{2n}) \phi_{\pi(n)}(y_{1n},y_{2n}){\rm d}y_{1n}{\rm d}y_{2n}
		\\& = \int_{((a,b)\times (c,d))^{n}}\sum_{\pi \in \sigma_n}(-1)^\pi K(x_{11},x_{21},y_{1\pi(1)},y_{2\pi(1)})\cdots K(x_{1n},x_{2n},y_{1\pi(n)},y_{2\pi(n)}) \\&~~~~ \phi_1(y_{11},y_{21}) \cdots \phi_n(y_{1n},y_{2n}) {\rm d}y_{11} \cdots {\rm d}y_{2n}. 
	\end{aligned}
\end{equation}
\fi
Therefore $Q_n \Lambda^n(T) Q_n$ is an integral operator with $2n$ variables, the kernel of which is 
\ifx\onecol\undefined
\begin{equation}
	\begin{aligned}
K({\bf x},{\bf y}) = &\sum_{\pi \in \sigma_n}(-1)^\pi K(x_{11},x_{21},y_{1\pi(1)},y_{2\pi(1)})\cdots \\&K(x_{1n},x_{2n},y_{1\pi(n)},y_{2\pi(n)}).
\end{aligned}
\end{equation}
\else
\begin{equation}
	\begin{aligned}
K({\bf x},{\bf y}) = \sum_{\pi \in \sigma_n}(-1)^\pi K(x_{11},x_{21},y_{1\pi(1)},y_{2\pi(1)})\cdots K(x_{1n},x_{2n},y_{1\pi(n)},y_{2\pi(n)}).
\end{aligned}
\end{equation}
\fi
Next we will show that the kernel is $L^2$ positive definite, which corresponds to $L^2PD(X,\mu)$ in \cite{ferreira2009eigenvalues}, where $X$ is the space $((a,b)\times (c,d))^{n}$ and $\mu$ is the Lebesgue measure. That is to say, we need to show $\int_X \left( \int_X K({\bf x},{\bf y})f({\bf y}) {\rm d} \mu({\bf y})\right)f^{*}({\bf x}) {\rm d} \mu({\bf x}) \geqslant 0$. First we can easily observe that for arbitrary $x_{1i},y_{1i} \in (a,b)$ and $x_{2i},y_{2i} \in (c,d)$, $K({\bf x},{\bf y})\geqslant 0$. The reason is that for given $x_{1i},y_{1i},x_{2i},y_{2i}$, $K({\bf x},{\bf y})$ is in fact the determinant of a correlation matrix, where the element in the $i_{\rm th}$ row and $j_{\rm th}$ column of the matrix is $K(x_{1i},x_{2i},y_{1j},y_{2j}) = \mathbb{E}[E(x_{1i},x_{2i})E^{*}(y_{1j},y_{2j})]$. The positive semidefinite property of a correlation matrix is easy to prove, which leads to the conclusion that any order principal minor determinant of it is non-negative. Therefore, $K({\bf x},{\bf y})$ is non-negative for any given ${\bf x}$ and ${\bf y}$ in the region $X$. From \cite[{\bf Theorem 2.1}]{ferreira2009eigenvalues} we know that the kernel is $L^2PD(X,\mu)$. Then, according to \cite[{\bf Theorem 2.4}]{ferreira2009eigenvalues} we know that the trace of the integral operator $Q_n \Lambda^n(T) Q_n$ equals $\int_{((a,b)\times (c,d))^{n}} K({\bf x},{\bf x}) {\rm d}x_{11} \cdots {\rm d}x_{2n}$, which means that 
\ifx\onecol\undefined
	\begin{equation}
	\begin{aligned}
		&{\rm tr}\left( \Lambda^n ({T}) \right) = \frac{1}{n!} \int_{((a,b)\times (c,d))^{n}} \Bigg( \sum_{\pi \in \sigma_n}(-1)^\pi \\&~~~~~~~~\prod_k K(x_{1k},x_{2k},x_{1\pi(k)},x_{2\pi(k)})\Bigg){\rm d}x_{11}\cdots {\rm d}x_{2n}.
	\end{aligned}
	\label{equ_tr_T_2variable}
\end{equation}
\else
\begin{equation}
	\begin{aligned}
	{\rm tr}\left( \Lambda^n ({\bf T}) \right)& = \frac{1}{n!} \int_{((a,b)\times (c,d))^{n}} \Bigg( \sum_{\pi \in \sigma_n}(-1)^\pi \prod_k K(x_{1k},x_{2k},x_{1\pi(k)},x_{2\pi(k)})\Bigg){\rm d}x_{11}\cdots {\rm d}x_{2n}.
	\end{aligned}
	\label{equ_tr_T_2variable}
\end{equation}
\fi

By substituting (\ref{equ_tr_K_2variable}) and (\ref{equ_tr_T_2variable}) into (\ref{equ_exterior_algebra}), we can extend {\bf Lemma 2} to integral operator with two variables, which corresponds to the analysis of two-dimensional transceivers. The other parts of the convergence analysis and the extension to scenarios like three-dimensional transceivers have no essential difference from what we already have done in the paper, so we will omit these contents in this paper.

\ifCLASSOPTIONcaptionsoff
 \newpage
\fi

\footnotesize

\bibliographystyle{IEEEtran}

\bibliography{IEEEabrv,bib}

\end{document}